\documentclass[3p,preprint]{elsarticle}
\newcommand{\abbreviations}[1]{%
  \nonumnote{\textit{Abbreviations:\enspace}#1}}
\usepackage{graphicx,pdfpages,float}
\usepackage{mathtools}
\usepackage[utf8x]{inputenc}
\usepackage[pdfencoding=auto]{hyperref}
\hypersetup{%
bookmarksnumbered=true,%
colorlinks=true,%
pdfkeywords={Timing System, RF Synchronization, Bucket Selection, Pulse to Pulse Modulation, SuperKEKB, AC Drift}
}

\begin{document}

\begin{frontmatter}

\title{Analysis and Stabilization of AC Line Synchronized Timing System for SuperKEKB}
\author[1,2]{Di Wang\corref{cor1}}\ead{sdcswd@post.kek.jp}
\author[1,2]{Kazuro Furukawa}\ead{kazuro.furukawa@kek.jp}
\author[1,2]{Masanori Satoh}\ead{masanori@post.kek.jp}
\author[1]{Hiroshi Kaji}
\author[1]{Hitoshi Sugimura}
\author[1]{Yoshinori Enomoto}
\author[1]{Fusashi Miyahara}
\cortext[cor1]{Corresponding author}
\address[1]{High Energy Accelerator Research Organization (KEK), 1-1 Oho, Tsukuba, Ibaraki, 305-0801, Japan}
\address[2]{The Graduate University for Advanced Studies, Hayama, Kanagawa, 240-0193, Japan}
\abbreviations{\textbf{PPM}, Pulse-to-Pulse Modulation; \textbf{EVG}, Event Generator; \textbf{EVR}, Event Receiver; \textbf{BSC}, Bucket Selection Cycle; \textbf{MTG}, Master Trigger Generator; \textbf{TDC}, Time-to-Digital Converter.}

\hyphenation{SuperKEKB}

\begin{abstract}
A timing system provides high-precision signals to allow the controls over a variety of hardware and software components in the accelerator complex. This is guaranteed by the radio frequency (RF) and trigger signal synchronization for subsystems such as klystrons, pulsed magnets, and beam monitors. The main trigger signal should be distributed throughout the facility and repeated at the beam repetition rate. This trigger signal is usually generated by the same phase of an AC power line to follow the source of the fluctuation of an electrical grid and reduce the unwanted variation of the beam quality. To fulfill the needs of the multi-accelerator facility at KEK, apart from the normal trigger synchronization and bucket selection injection control, a beam operation scheme called the pulse-to-pulse modulation is utilized; hence, the complexity of the timing system increases. Uncertainty in the system and a trigger signal delivery error caused by a drastic AC power line drift are observed. Further, the effort to establish a reliable timing system at KEK and several solutions to improve the system robustness are presented.

\end{abstract}


\end{frontmatter}

\section{Introduction}
\label{sec:introduction}

\begin{table*}[!h]
\begin{center}
\caption{Particle and parameters of LINAC beam for four destinations.}
\begin{tabular}{cccc}
\hline 
\textbf{Direction}          & \textbf{Particle}     & \textbf{Energy~(GeV)}     & \textbf{Charge~(nC)}\\
\hline
 SuperKEKB-DR   & $e^+$         & 1.1   &   4.0\\
 SuperKEKB-LER  & $e^+$         & 4.0   &   4.0\\
 SuperKEKB-HER  & $e^-$         & 7.0   &   4.0\\
 PF     & $e^-$         & 2.5   &   0.2\\
 PF-AR  & $e^-$         & 6.5   &   0.2\\
\hline
\end{tabular}
\label{table:beam_type}
\end{center}
\end{table*}

\begin{figure}[!hbt]
\centering
\includegraphics*[width=.8\columnwidth]{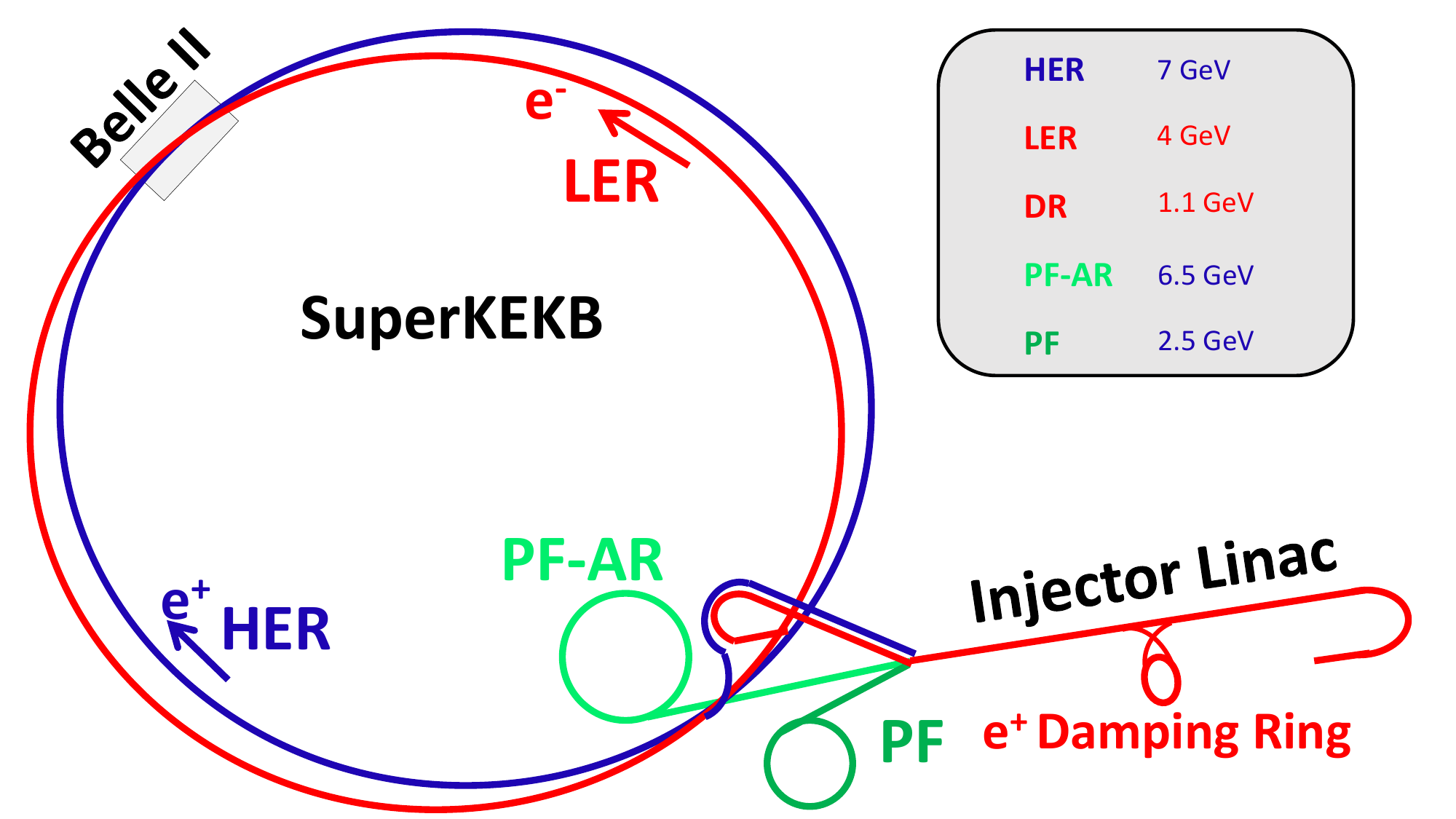}
\caption{Overview of LINAC, SuperKEKB, and PF/PF-AR.}
\label{fig:skb_layout}
\end{figure}

SuperKEKB is an asymmetric-energy electron-positron double-ring collider whose scientific target is to make significant advancements at the luminosity frontier. The SuperKEKB collider complex consists of a 700-m long injector linear accelerator (LINAC) and two main rings (MRs), which include a 7-GeV high-energy electron ring (HER) and a 4-GeV low-energy positron ring (LER). The 1.1-GeV positron damping ring (DR), which is used to reduce the positron beam emittance, is in the middle of LINAC. Apart from supplying the beam to the SuperKEKB collider, LINAC also provides beams for two synchrotron light source facilities, i.e., the 2.5-GeV PF and 6.5-GeV PF-AR. The machine parameters are listed in Table~\ref{table:beam_type}. A schematic view of LINAC and the rings are shown in Fig.~\ref{fig:skb_layout}. To modulate the beam properties, a methodology called pulse-to-pulse modulation (PPM) is first utilized at CERN to fulfill the prerequisites for multi-accelerator facilities~\cite{1977thepsstaffPulsePulseModulation}. This ideology is also implemented at KEK LINAC to share LINAC among four rings and improve its injection efficiency~\cite{2010furukawaPulsetopulseBeamModulation}.

The existing accelerator timing system of LINAC is responsible for timing signal generation, RF synchronization, and bucket selection. A stable timing signal for beam control devices is required to operate at a scheduled sequence. Under the control of high-resolution timing signals, the power supply of pulsed magnets is triggered to maintain a stable magnetic field\cite{2018enomotoPulsetopulseBeamModulation,2020enomotoPulsedMagnetControl}. The downstream RF system must operate at an appropriate phase to achieve the desired accelerating effect for the bunch. Finally, the bunch injects into the target ring within an allowable timing jitter with the help of an injection pulsed septum magnet and kicker magnet. This process repeats at a beam repetition rate of 50 Hz at LINAC.

At the injection point, the RF phase between LINAC and MRs must be synchronized; otherwise, the bunch gradually becomes lost owing to the longitudinal synchrotron oscillation. Consequently, the injection opportunity arises based on the common frequency between LINAC and MR RF frequency.

To equalize the bunch distribution inside the ring, the injection RF bucket should be selected based on the timing system for every pulse. This procedure is called bucket selection~\cite{2015kajiBucketSelectionSystem}. The bucket selection is implemented by introducing some delay, which is calculated by converting the RF bucket number to the corresponding delay time, and subsequently, to the timing signal.


The timing signal is often the key to a stable beam operation because many of the beam properties are dependent on the timing of the device operation. The bunch arrival timing error or abnormal bunch energy generated by the timing control system is likely to trigger the beam abort, which is used to protect the Belle II detector, to divert all circulated beams into a beam dump~\cite{2014mimashiSuperKEKBBeamAbort}. Consequently, the beam abort affects the integrated luminosity accumulation goal of SuperKEKB~\cite{2013ohnishiAcceleratorDesignSuperKEKB}. At SuperKEKB, to maintain a small aperture at the collision point, the MR injection timing jitter should be less than 30 ps (r.m.s)~\cite{2003furukawaTimingSystemKEKB,2009furukawaEventBasedTimingControl,1995KEKBBFactoryDesign} and this is guaranteed through RF synchronization between the LINAC, DR, and MR. This requirement can be relaxed to 300 ps (r.m.s) for PF and PF-AR~\cite{2009furukawaEventBasedTimingControl}.

\section{Overview of timing system at LINAC}
\label{sec:ts_linac}
\subsection{RF synchronization}
\begin{figure}[!hbt]
\centering
\includegraphics*[width=.85\columnwidth]{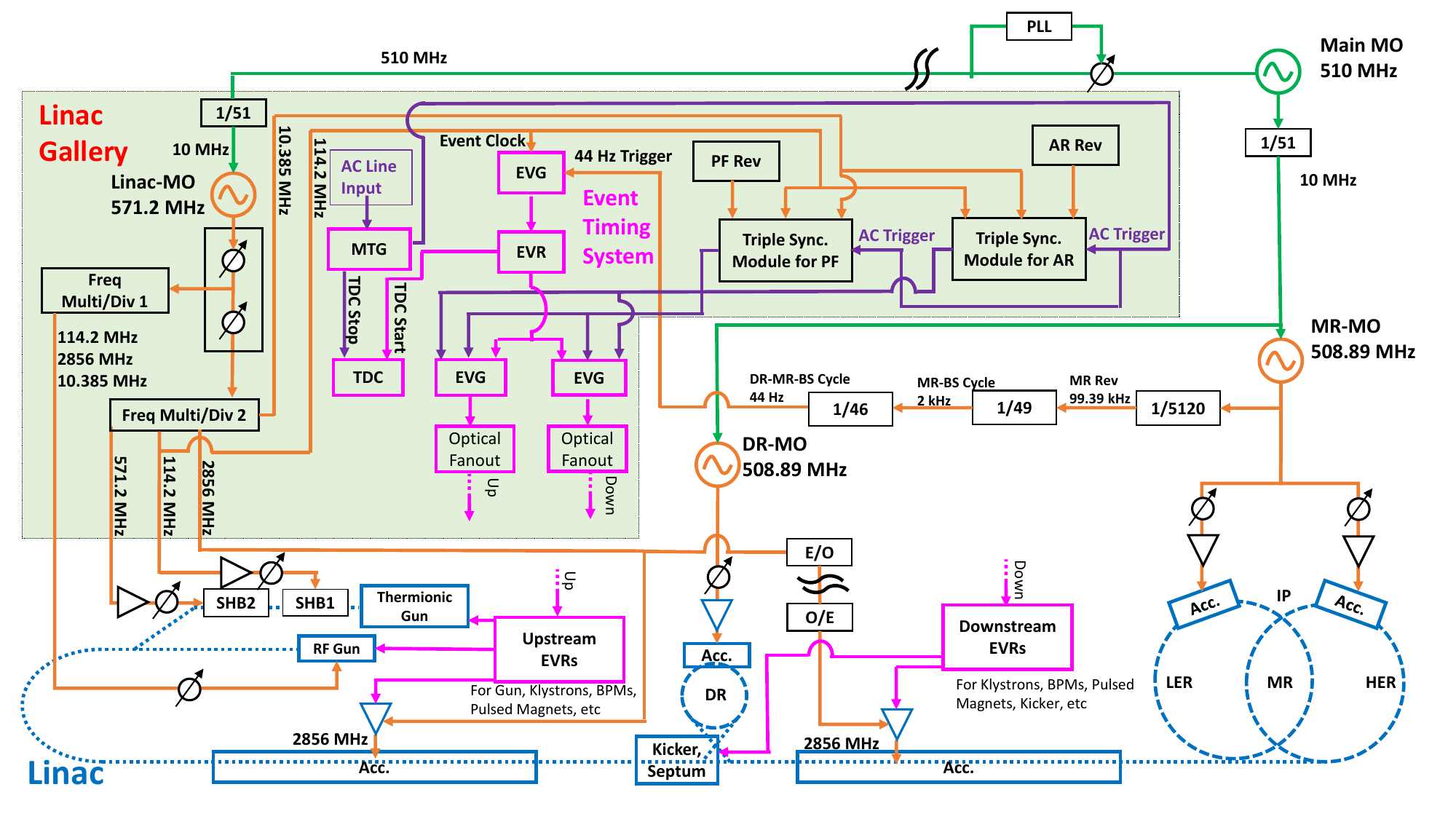}
\caption{Block diagram of driver and event timing system for LINAC, DR, and MR. Rev indicates the revolution frequency, BS indicates the bucket selection, E/O indicates the electrical/optical converter, and Acc. Indicates the accelerator structure}
\label{fig:linac_trigger}
\end{figure}

Figure~\ref{fig:linac_trigger} depicts the RF frequency relation of the LINAC timing system. The LINAC master oscillator (LINAC-MO) is synchronized to the MR master oscillator (MR-MO) and DR master oscillator (DR-MO) by the external 10-MHz reference generated by a frequency division of the 510-MHz main master oscillator (MMO). The phase drift caused by the long-distance transmission is compensated by a phase lock loop (PLL) circuit. The frequencies of LINAC-MO and MR-MO are 571.2 and 508.89 MHz, respectively. The DR-MO follows the MR frequency. The largest common frequency (CF) between LINAC-MO and MR-MO is 10.385 MHz, which means that 55 cycles of LINAC-MO cover exactly 49 cycles of MR-MO. The 11th harmonic (114.24 MHz) and 55th harmonic (571.2 MHz) of the CF are used to drive two subharmonic bunchers (SHB) located after the thermionic gun. The 2856-MHz RF signal is used to drive the accelerator structures. The timing system also uses the 114.24-MHz signal as the event clock to deliver the event codes. A detailed description of the event timing system is discussed in Section~\ref{sub:event_timing_system}.

For PF and PF-AR, the ring RF frequency is independent of the LINAC RF, and a specific algorithm is implemented to acquire the coincidence signal between the LINAC RF and PF/PF-AR RF frequencies. The details are described in~\cite{2019miyaharaTimingSystemMultiple}.

\subsection{Event-based timing system}
\label{sub:event_timing_system}

Two concepts for designing timing systems are widely used in accelerator installations. One is to supply timestamps to the control devices and correlate the timestamp over the timestamp-based timing network simultaneously~\cite{2011serranoAcceleratorTimingSystems}. Another common solution is the event-based timing system that generates and receives event codes from the event generator (EVG) and the event receiver (EVR), respectively. The EVG module receives an RF as the event clock to transfer the data stream through the optical fibers to EVRs. The data stream includes the 8-bit event code and another 8-bit data which is shared by the distributed bus and data buffer~\cite{VMEEVG230TechnicalReference}. The 8-bit event codes with attached delivery delay times are stored in the sequence RAM of EVG and sent out to EVRs when the trigger signal arrives. The source of the trigger signal can be configured as an external signal, an AC power line synchronization logic, or the internal multiplexed counter of the EVG. Some other modules, such as a fan-out module and universal I/O modules, are usually needed to build the timing network. Upon receiving the desired event code, the EVR can generate VME interrupt and a trigger signal with specified delay and width to equipment, such as a pulsed magnet or BPM.

\begin{figure}[!hbt]
    \centering
    \includegraphics*[width=.8\columnwidth]{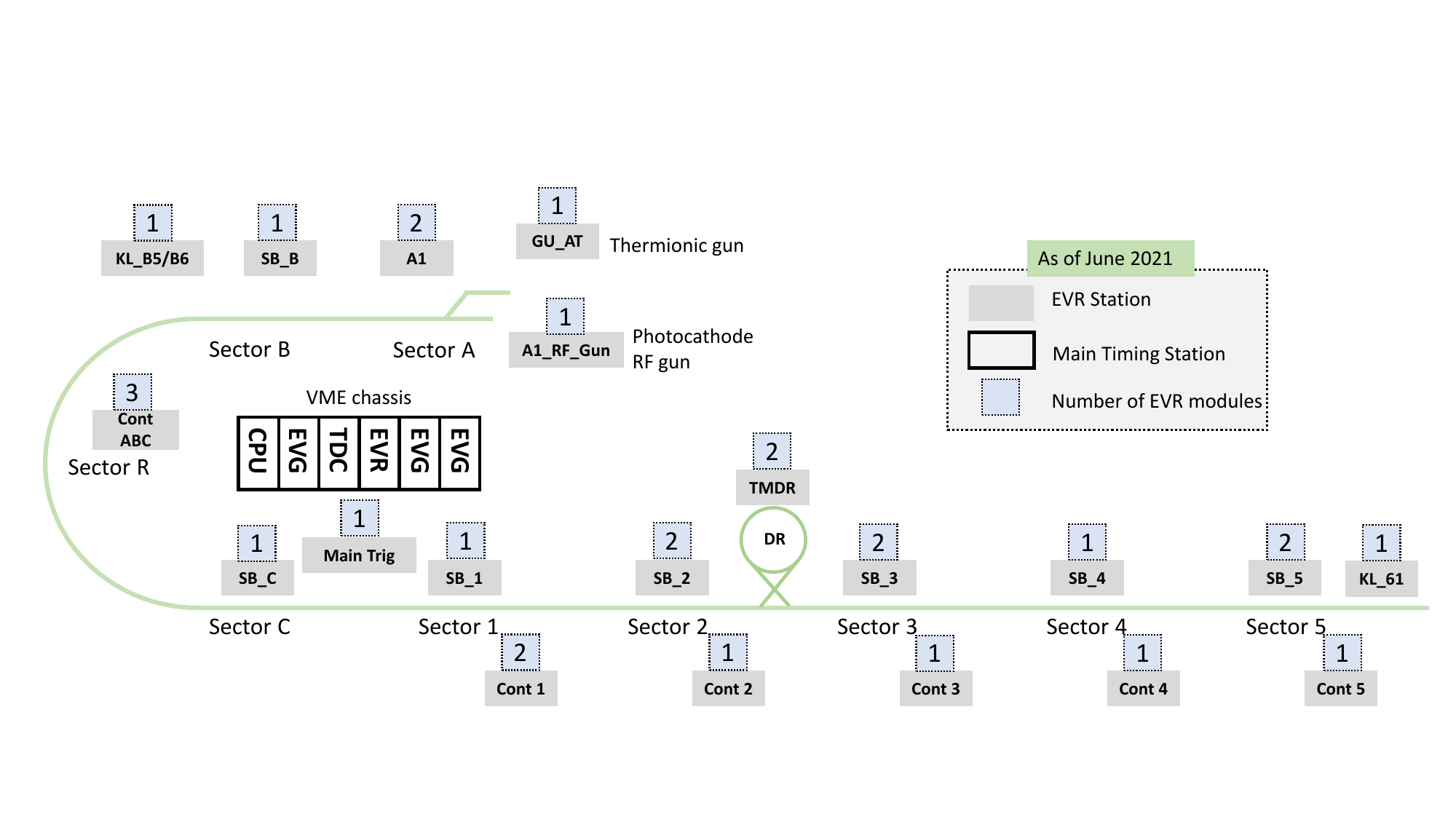}
    \caption{Overall configuration of the event-based timing stations at LINAC.}
    \label{fig:simple_event_system}
\end{figure}

The event timing system was introduced at KEK LINAC in 2008~\cite{2008furukawaDevelopmentTimingControl,2009furukawaNewEventbasedControl} and the commercial VME-based event modules supplied by the MRF company and SINAP are used at LINAC~\cite{MicroResearchFinland, 2015kajiNewEventTiming}. A schematic view of the event timing system currently used at LINAC is shown in Fig.~\ref{fig:simple_event_system}. Three MRF-VME-EVG230 modules and an MVME6100 CPU module are installed in the main timing station~\cite{2015kajiInstallationCommissioningNew}. The main timing station delivers events to 20 local EVR stations at KEK LINAC. The event clock is chosen to be 114.24 MHz and the trigger signal comes from the 50-Hz AC power line (i.e., AC50). More than 120 event codes are defined to modulate the beam properties. The distributed bus and data buffer functions are also used to deliver beam gate information~\cite{2018kajiBeamGateControl} and injection parameters, such as bucket selection delay and RF phase~\cite{2018sugimuraSynchronizedTimingControl}.

\subsection{Bucket selection} 
\label{sub:bucket_selection}

For the ring accelerators, the ring RF $f_{rf}$ must satisfy
\begin{equation}
    f_{rf} = h*f_{rev}
\end{equation}
\noindent
\noindent
where $h$ is the harmonic number (i.e., the number of RF buckets), and $f_{rev}$ is the revolution frequency.

The primary task for bucket selection is to provide the ability to select an arbitrary RF bucket in the ring for a single injection pulse. This can be achieved by adding a proper delay time to the gun triggering signal after defining the fiducial bucket in the ring. As the stable phase of RF cavity coincides based on the common frequency (CF) between LINAC and MR, the injection is performed based on the period of the CF~\cite{2015kajiBucketSelectionSystem}. After $h$ times injection, all RF buckets are filled. The period during which all ring RF buckets can be filled is defined as a bucket selection cycle (BSC) and the period of BSC can be represented as
\begin{equation}
    T_{BSC} = h_{MR} * {T_{CF}}
\end{equation}
\noindent
\noindent
where $T_{CF}$ is the period of common frequency between LINAC and MR.

\begin{table*}[!h]
    \begin{center}
    \caption{Bucket selection parameters for DR and MR.}
    \begin{tabular}{cccc}
    \hline
        & \textbf{DR}       & \textbf{MR} & \textbf{Unit}\\
    \hline
     RF frequency   & 508.89        & 508.89 & MHz\\
     Circumference  & 135.5         & 3016.3 & m\\
     Revolution frequency   & 2.21  &  0.099 & MHz\\
     RF buckets     & 230   & 5120 & -\\
     BSC frequency      & 45.2  & 2.03 & kHz \\
     BSC period     & 22.1  & 493 & $\mu$s \\
    \hline
    \end{tabular}
    \label{table:bs_param}
    \end{center}
\end{table*}

For example, according to Table ~\ref{table:bs_param}, the harmonic number $h_{MR}$ at two MRs are 5120, and the injection opportunity arises at the common frequency of 10.385 MHz between LINAC and the SuperKEKB MR (i.e., every 96.3 ns); hence, the BSC for MR, $T_{BSC}$, can be calculated as
\begin{equation}
    T_{BSC} = 5120*96.3\ ns = 493\ \mu s
\end{equation}
\noindent
\noindent
which means all RF buckets can be selected within 493 $\mu$s, and the bucket selection process can be easily accomplished every 20 ms.

Nevertheless, if DR is considered, the complexity of the bucket selection system grows. The harmonic number at DR is 230 and the least common multiple between the DR and LER harmonic numbers is 117760; thus, the BSC for DR and LER becomes 11.34 ms (i.e., 88 Hz). Table~\ref{table:bs_delay} lists the relation of the example delay time and ring RF bucket number.

\begin{table*}[!hbt]
\begin{center}
\caption{The relation between injection delay timing and bucket numbers at DR and MR.}

\begin{tabular}{lccc}
\hline
\textbf{Opportunity}    & \textbf{Delay}        & \textbf{DR Bucket}    & \textbf{MR Bucket}\\
\hline
 0      & 0 ns          & 0     & 0\\
 1      & 96.3 ns       & 49    & 49\\
 2      & 192.6 ns      & 98    & 98\\
 3      & 288.9 ns      & 147   & 147\\
 ..     & ..        & ..    & ..\\
 230    & 22.1 $\mu$s & 0   & 1030\\
 ..     & ..        & ..    & ..\\
 5120   & 493 $\mu$s        & 180   & 0\\
 ..     & ..        & ..    & ..\\
 20771  & 1.99 ms       & 29    & 4019\\
 20772  & 2 ms          & 78    & 4068\\
 ..     & ..        & ..    & ..\\
 117760     & 11.34 ms      & 0     & 0\\
 \hline
\end{tabular}
\label{table:bs_delay}
\end{center}
\end{table*}

\subsection{AC line trigger} 
\label{sub:ac_sync}
As previously stated, the event timing triggering signal is generated at the same AC50 phase. It is usually necessary to maintain the beam intensity and quality at the same level for consecutive triggers. The experiment done in the KEKB era showed that the beam quality deterioration, such as beam energy jitter, has a strong dependency on the triggering phase of the AC-powered filament inside klystron~\cite{2021kazuroPrivateCommunication}. Therefore, the beam timing was triggered at a fixed phase of AC50 for the KEKB timing system~\cite{2003furukawaTimingSystemKEKB,2004suwadaTriggertimingSignalDistribution}. This requirement was also kept for the SuperKEKB timing system.

The frequency of the AC line is not ideal 50 Hz. The Tokyo Electric Power Company, which supplies AC power for KEK, adjusts the frequency of the AC power line within 50$\pm0.2$ Hz to balance the supply and demand requirement of the electricity market~\cite{2020RuleFrequencyAdjustment}. The timing system is required to follow the drift of AC50.

For HER injection, the BSC is 493 $\mu$s and the gun timing can be calculated based on the coincidence between AC50 and BSC. However, for LER injection with DR, the BSC is longer and to be 11.34 ms, so it cannot coincide with AC50 every 20 ms. Therefore, a method called sequence shift is utilized to deliver the triggers at the same phase of AC50, while it is also synchronized with the BSC.


\subsection{Sequence shift}
\label{sub:seq_shift}

The implementation of sequence shift requires several modules to provide adequate information to meet the requirement of AC line trigger. A Master Trigger Generator (MTG) module with a voltage comparator circuit was manufactured to generate the AC50 signal at the fixed phase from the AC power line. A time-to-digital converter (TDC), which receives the ``TDC start'' and ``TDC stop'' signal, is used to measure the AC50 delay time with a resolution of 1 ns~\cite{2015suwadaWideDynamicRange}.

\begin{figure}[!hbt]
    \centering
    \includegraphics*[width=.9\columnwidth, height=8cm]{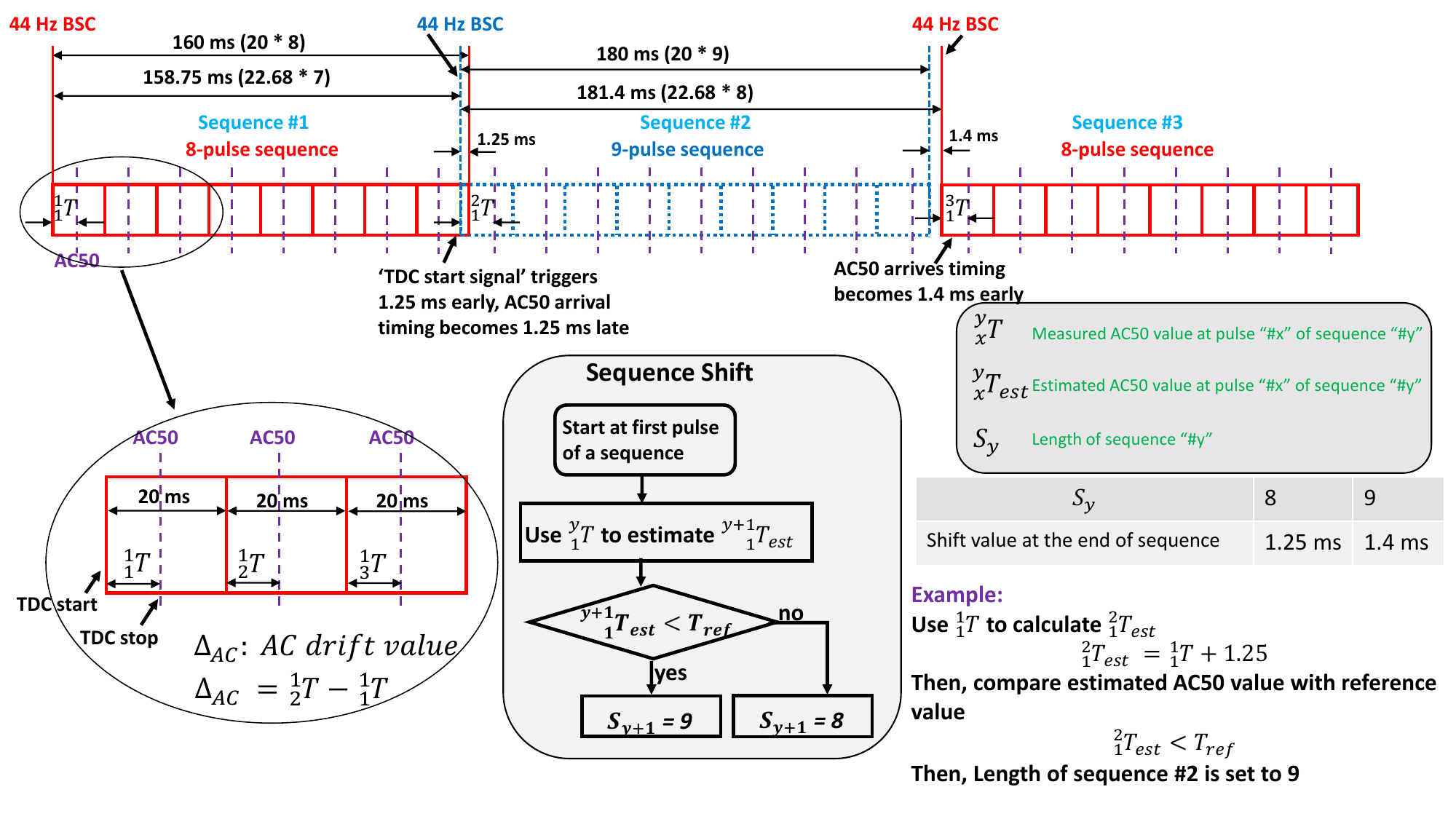}
    \caption{The algorithm of 8/9-pulse sequence shift. The AC50 arrival time in current pulse is used to estimate the AC50 arrival time in the future pulse.}
    \label{fig:sequence_shift_logic}
\end{figure}

The schematic view of 8/9-pulse sequence shift is shown in Fig.~\ref{fig:sequence_shift_logic}. The sequence length was selected to 8 pulses and 9 pulses in 2015~\cite{2015kajiInstallationCommissioningNew}. Inside the 8/9-pulse sequence, the requirements for synchronizing to both 44 Hz BSC and AC50 are fulfilled at every pulse. The timing system starts with the 44 Hz BSC which serves as the fiducial point of bucket selection. Note that 44 Hz (i.e., 22.68 ms) rather than 88 Hz BSC is used because it is close to 20 ms and simplifies the system design. The pulse length is ideal 20 ms and different beam properties are modulated pulse-to-pulse. The AC50 arrival delay at every pulse is measured by the TDC module and is used for generating AC50-synchronized beam timing signal. The range of AC50 arrival delay values is from 0 to 20 ms. The time distance between the fiducial point of bucket selection and AC50 in every pulse can be easily calculated based on the pulse number. For example, if the AC50 arrival delay measured in the third pulse is 10 ms, the overall delay relative to BSC is 50 ms. Then, the DR and LER bucket number can be acquired through the formulas described in~\cite{2018kajiInjectionControlSystem}.

As shown in Fig.~\ref{fig:sequence_shift_logic}, after an 8-pulse sequence, the discrepancy between ideal 20 ms pulse and 44 Hz BSC is 1.25 ms. The sequence ``\#2'' is launched 1.25 ms earlier to resynchronize with 44 Hz BSC. Accordingly, the AC50 arrival delay at the first pulse of sequence ``\#2'' becomes 1.25 ms later because the ``TDC start'' signal comes 1.25 ms earlier. Similarly, the sequence ``\#3'' is launched 1.4 ms later after a 9-pulse sequence.

As the AC50 always drifts, the AC50 arrival timing can be out of the boundary in a pulse and break the synchronization requirement. The sequence shift algorithm is then used for compensating the AC50 drift and keeping AC50 arrival timing appear in the middle of a pulse. To explain the sequence shift algorithm, let $\prescript{y}{x}{T}$ denote the AC50 arrival delay which is measured by TDC at pulse ``\#x'' of sequence ``\#y', $\prescript{y}{x}{T}_{est}$ denote the estimated AC50 value at pulse ``\#x'' of sequence ``\#y', $S_{y}$ denote length of sequence ``\#y'' and $T_{ref}$ denote the reference value for AC50 arrival delay. The value of $S_{y}$ is either 8 or 9 which corresponds to AC50 shift value of 1.25 ms and 1.4 ms, respectively. The value of $T_{ref}$ is usually selected to be around 10 ms. The AC50 drift value $\Delta_{AC}$ can be expressed by the discrepancy of the AC50 arrival delay values for two continuous pulses.

The main purpose of sequence shift algorithm is to decide the length of next sequence and keep $\prescript{y}{x}{T}$ be close to $T_{ref}$. For example, the type of sequence ``\#2'' is decided at the first pulse of sequence ``\#1''. The estimated AC50 arrival timing $\prescript{2}{1}{T}_{est}$ is derived from the TDC measurement value $\prescript{1}{1}{T}$ and sequence shift value (i.e., 1.25 ms). If $\prescript{2}{1}{T}_{est}$ is smaller than $T_{ref}$, the type of next sequence is selected as 9-pulse sequence; otherwise, 8-pulse sequence is selected.

\begin{figure}[!hbt]
    \centering
    \includegraphics*[width=.9\columnwidth]{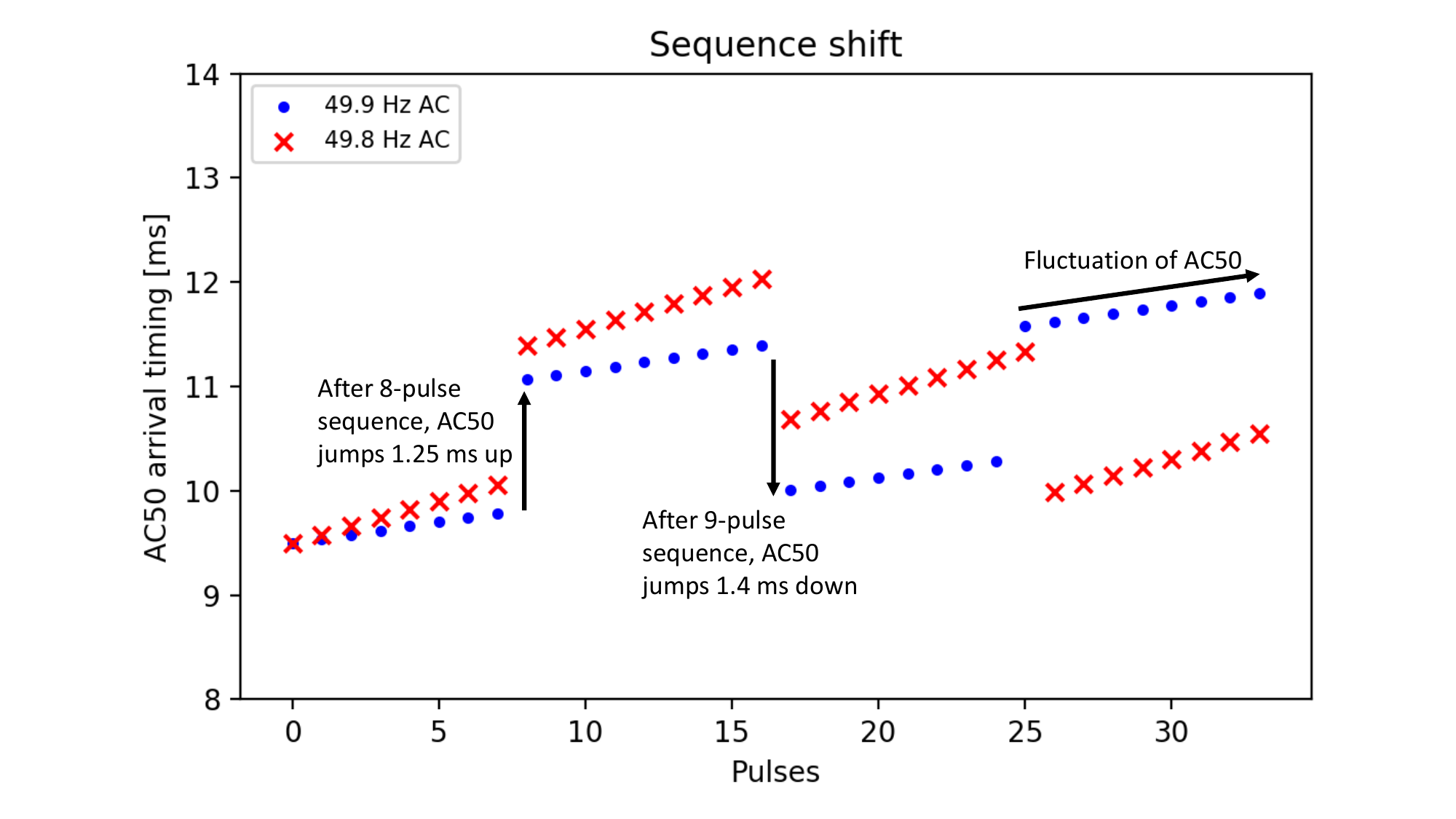}
    \caption{Example of sequence shift under two kinds of AC50 frequencies. The sequence shift algorithm makes different decisions for AC line frequencies of 49.8 Hz and 49.9 Hz. }
    \label{fig:sequence_example}
\end{figure}

Figure~\ref{fig:sequence_example} illustrates the example sequence shift under different AC50 frequencies. Every point represents the measured AC50 arrival delay in an injection pulse. The AC50 drift value $\Delta_{AC}$ is 40 $\mu$s and 80 $\mu$s for the AC line frequency of 49.9 and 49.8 Hz, respectively. The initial AC50 arrival timing is 9.5 ms and the AC50 reference value $T_{ref}$ is 9.85 ms. Two different decisions on the length of the next sequence are made at the first pulse of the second sequence. Larger AC50 arrival timing caused by fluctuation of AC50 is corrected by sequence shift. Consequently, the AC50 arrival delay value oscillates around the $T_{ref}$. The long-term stability of the 8/9-pulse sequence shift was studied in 2015 and loss of triggers was not observed~\cite{2015kajiInstallationCommissioningNew}.

The DR operation is carried in 2018. To achieve a satisfactory damping effect, a DR storage time of at least 40 ms is required. There also exists a restriction of the maximum storage time in the DR. The injection and extraction delay are calculated together, and both follow the AC line phase at the injection pulse. Thus, the extraction timing trigger has a larger discrepancy at the downstream LINAC for a long storage time. The difference between the extraction timing and AC50 arrival timing at extraction pulse becomes larger when the storage time increases. Consequently, the maximum storage time is chosen to be 200 ms, while the minimum storage time is 40 ms. To supply the requirement of the maximum DR storage time (200 ms), the sequence length is doubled to apply the 16/18-pulse sequence shift in 2018, and the sequence shift values are enlarged to 2.5 and 2.8 ms~\cite{2018kajiInjectionControlSystem}.



\section{Structure of the main timing station at LINAC}

\begin{figure}[!hbt]
    \centering
    \includegraphics*[width=.9\columnwidth, height=9cm]{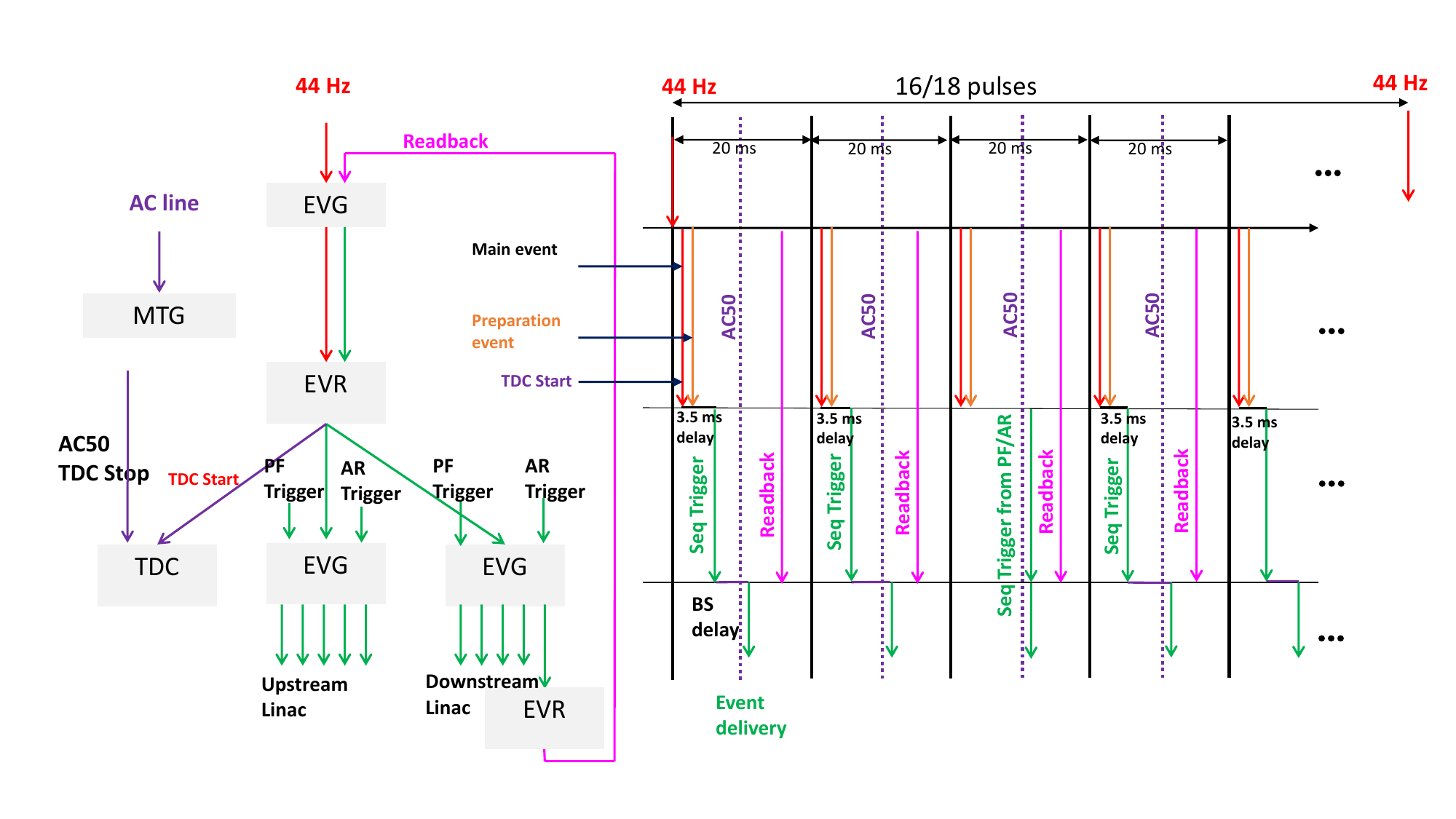}
    \caption{The timing chart of the main timing station at LINAC. Three EVGs are utilized to generate event codes and control the injection procedure. The upper-level EVG provides event codes to the middle-EVR, whereas the middle EVR triggers two lower-level EVGs and the two lower-level EVGs operate at 50 Hz. With the help of this architecture, the requirement that originates from the AC line synchronization is guaranteed.}
    \label{fig:event_timing}
\end{figure}

Figure~\ref{fig:event_timing} shows the structure of the main timing station at LINAC. Note that only the components that are related to the analysis of the AC50 dependency are illustrated here, whereas other parts, such as PF/PF-AR triple-sync logic, are omitted.

As there are four destination rings for LINAC, the main timing station needs to modulate the beam mode pulse-to-pulse. The principle of PPM is to modulate the beam properties of every pulse, by switching the control parameters, to improve the efficiency of LINAC. Every pulse corresponds to a beam mode. After one pulse, the injection beam mode gets switched, and, hundreds of parameters are changed accordingly. In total, 10 beam modes are currently used at LINAC, including 4 for normal beam operation and 6 for beam studies.

Three EVGs and two EVRs are utilized to generate event codes and control the injection procedure. The MTG and TDC modules provide the ability to measure the AC50 arrival timing at every pulse and supply the data for the decision-making of the AC50-synchronized trigger and sequence shift algorithm. Owing to the DR operation, some event codes are not identified before or after DR, e.g., the event codes for DR injection (extraction) septum and kicker. Therefore, LINAC is logically separated into two parts: the upstream LINAC and the downstream LINAC.

The upper-level EVG is used for sequence shift. It is functioned with the 16/18-pulse sequence and triggered by the 44 Hz BSC. The BSC synchronizes with the event clock (114.24 MHz). The events of 16/18 pulses are scheduled and stored in the sequence RAM of upper-level EVG. The time spacing for these pulses is carefully set to 20 ms with the help of the event clock counter inside EVG. Every 20 ms, the main and preparation event are delivered to the middle-level EVR. After 320 or 360 ms, the upper-level EVG resynchronizes with BSC. The logic is shown in Fig.~\ref{fig:sequence_shift_logic}.

The middle-level EVR is responsible for generating output pulses from the main and preparation event. The output pulse, which is generated upon the arrival of the main event, triggers both the TDC and two lower-level EVGs. The triggering signal for TDC starts the TDC measurement immediately, whereas triggering for two lower-level EVG happens after 3.5 ms. This 3.5-ms delay, which is mainly used to reserve the runtime of the beam parameters switching, is achieved by adding the delay value of the pulse generator in EVR. Note that this 3.5-ms delay is determined by our operational experience and can be changed to other values. The preparation event is used to perform PPM.

The event codes and bucket selection delay values are set into the sequence RAMs of two lower-level EVGs every pulse. These events trigger devices at upstream LINAC and downstream LINAC, respectively.

A local EVR receives the event codes generated from two lower-level EVGs and sends a ``readback'' signal to the upper-level EVG to assure that the current event codes are successfully delivered. This ``readback'' signal also triggers the processing of bucket selection calculation of the next pulse.

The AC50 should always arrive in the middle of the 20 ms pulse to maintain a stable operation, as some data processing procedures of PPM, such as switching the beam mode and bucket selection delay calculation, are performed at the beginning and end of every pulse. This target is accomplished via sequence shift.

\section{Timing system failures}
\label{sec:timing_failures}

\subsection{Failure modes}
The timing system, which satisfies the requirements to operate the DR for the positron injection, was installed and commissioned in 2018~\cite{2018kajiInjectionControlSystem}. As the growth of the system complexity comes from the bucket selection and 16/18-pulse sequence shift, several failure modes of the timing system are observed. An event code log system, which monitors and saves all event codes received in EVR, is developed to diagnose the failure modes in 2019~\cite{2019wangFaultDiagnosisEvent}. The severity of some timing system failures is minimal because the trigger signals for devices are masked and inhibited by beam gate system~\cite{2018sugimuraTriggerControlSystem} when some abnormal events are transmitted. On the other hand, a timing system failure that stops the delivery of event codes for a few minutes is severe for the physics run because it usually triggers the beam abort system to dump all the beams at SuperKEKB. It normally takes 10 minutes for the operator to check the status of the devices and restart the injection. The occurrence of such kind of failure is rare (i.e., 9 times) in 2020. However, the number of failures increases to 18 times in a month since the 2021 spring, and the beam operation is frequently interrupted. Thus, it is urgent and significant to understand the failure cause and stabilize the operation. By analyzing the event code log system and auxiliary timing data, such malfunction of the timing system is caused by strong AC50 drift.

\subsection{Failure cause}

\begin{figure}[!hbt]
    \begin{minipage}[t]{0.5\linewidth}
        \centering
        \includegraphics[width=2.5in, height=5cm]{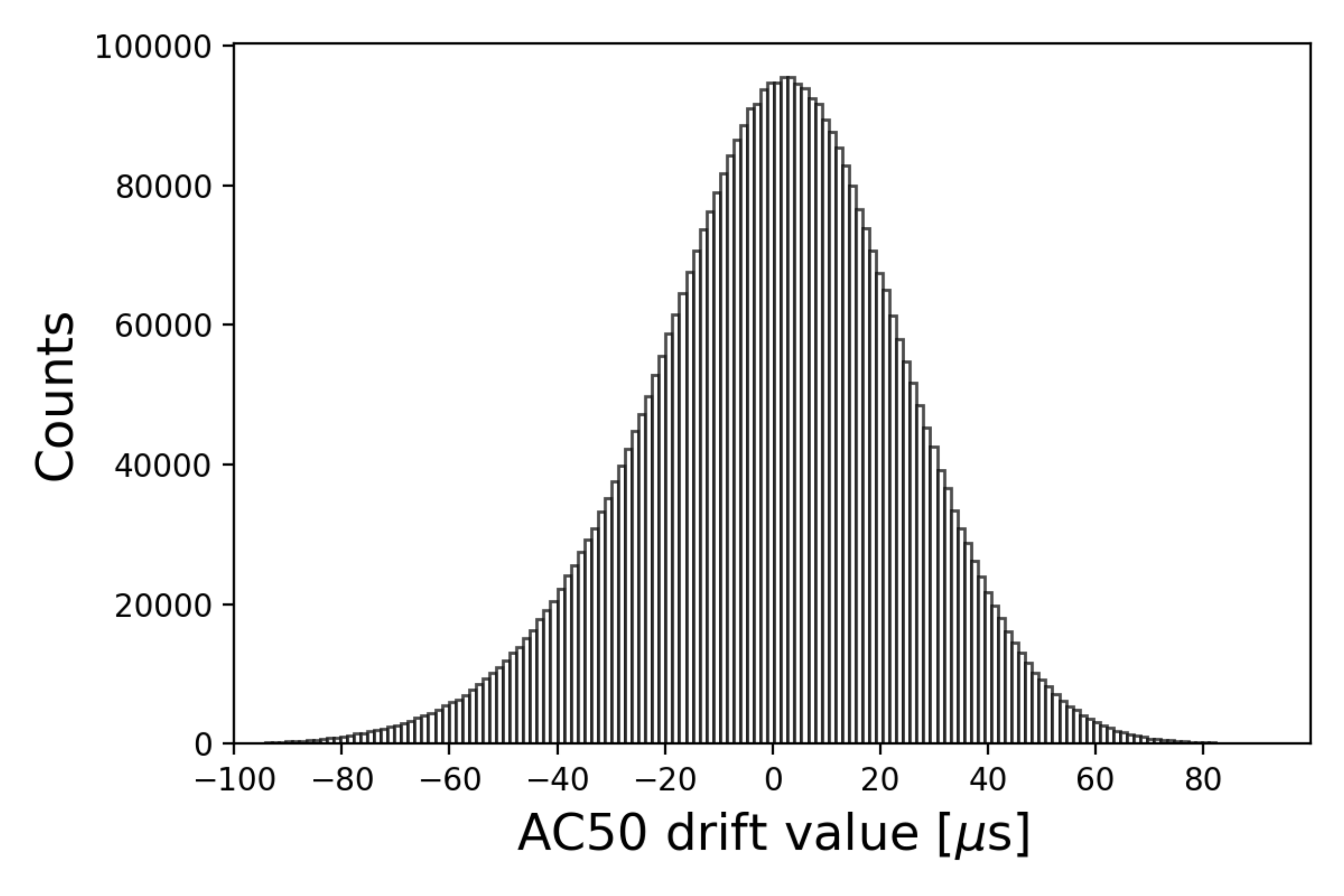}
        \caption{Histogram of AC50 drift value in 24 hours.}
        \label{fig:ac_drift_hist}
    \end{minipage}
    \begin{minipage}[t]{0.5\linewidth}
        \centering
        \includegraphics[width=2.5in, height=5cm]{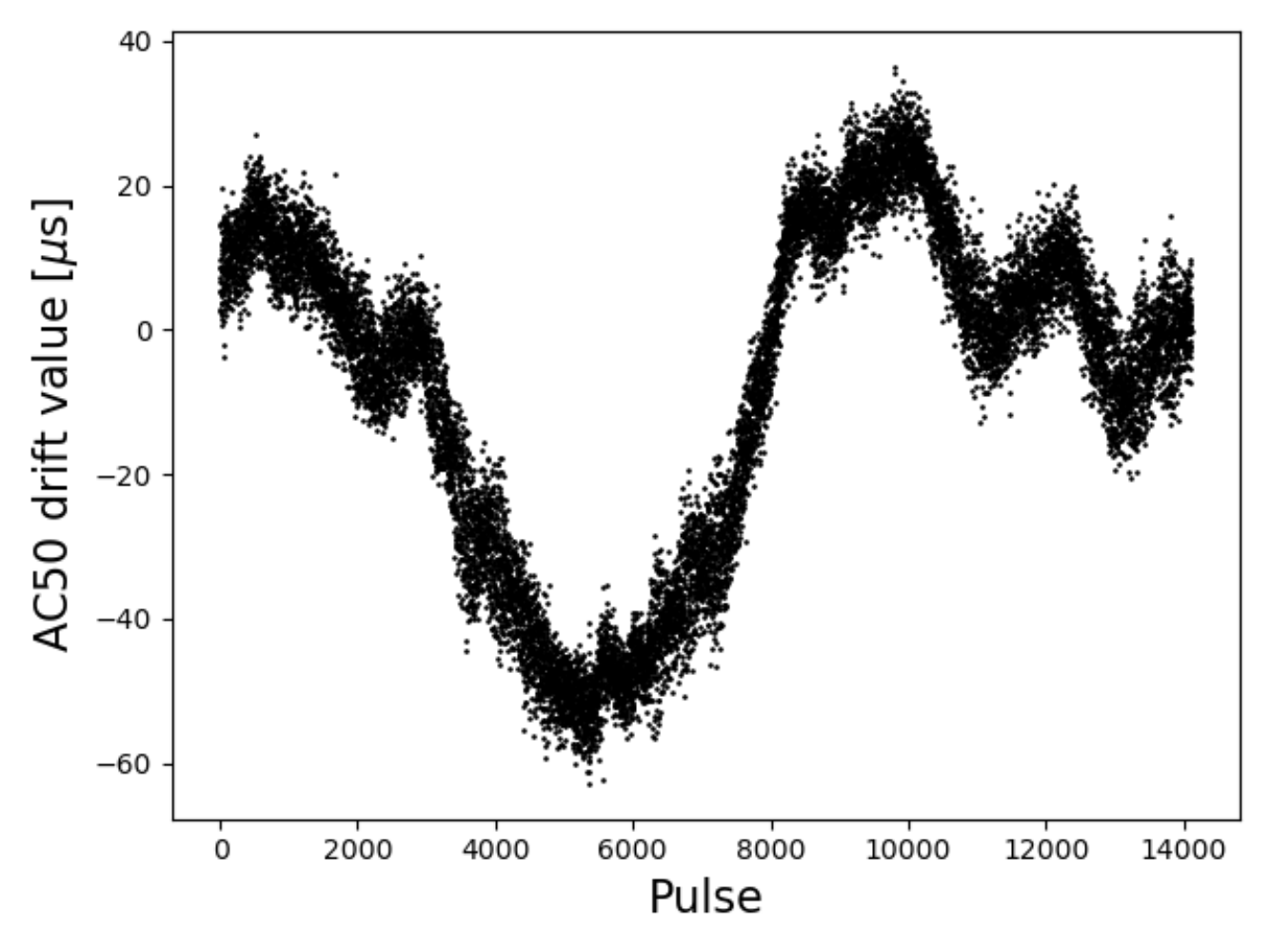}
        \caption{AC50 drift value in $\sim$5 minutes.}
        \label{fig:ac_drift_5_min}
    \end{minipage}
\end{figure}

The AC50 delivered from the electric power company always drifts. The histogram in Fig.~\ref{fig:ac_drift_hist} indicates the distribution of $\Delta_{AC}$ in 24 hours and the plot in Fig.~\ref{fig:ac_drift_5_min} displays the variation of $\Delta_{AC}$ in $\sim$5 minutes. The $\Delta_{AC}$ of about 99\% pulses are less than 40 $\mu$s. The maximal value of $\Delta_{AC}$ in one day is usually less than 100 $\mu$s. The 16/18-pulse injection sequence shift was designed to compensate for the AC50 drift. However, if AC50 continues drifting sharply, the balance is possibly disrupted and the AC50 arrival timing becomes away from $T_{ref}$. If the AC50 appears at the very beginning or end of the 20 ms pulse, the program processing logic can be interfered with by the neighboring pulses~\cite{2019wangFaultDiagnosisEvent}. If sufficient time for the preparation of the next injection cannot be ensured, the event codes might be sent out erroneously without finishing the preparation. The race condition for the timing system at LINAC is a special case when the AC50 arrival timing measured by TDC is smaller than 4.5 ms or larger than 15 ms. This threshold of the race condition is determined based on the operational experience. After entering the race condition, the event code delivery is highly probable to stop, and the beam abort at SuperKEKB MR is triggered when the timing system fails.

\begin{figure}[!hbt]
    \centering
    \includegraphics*[width=.8\columnwidth]{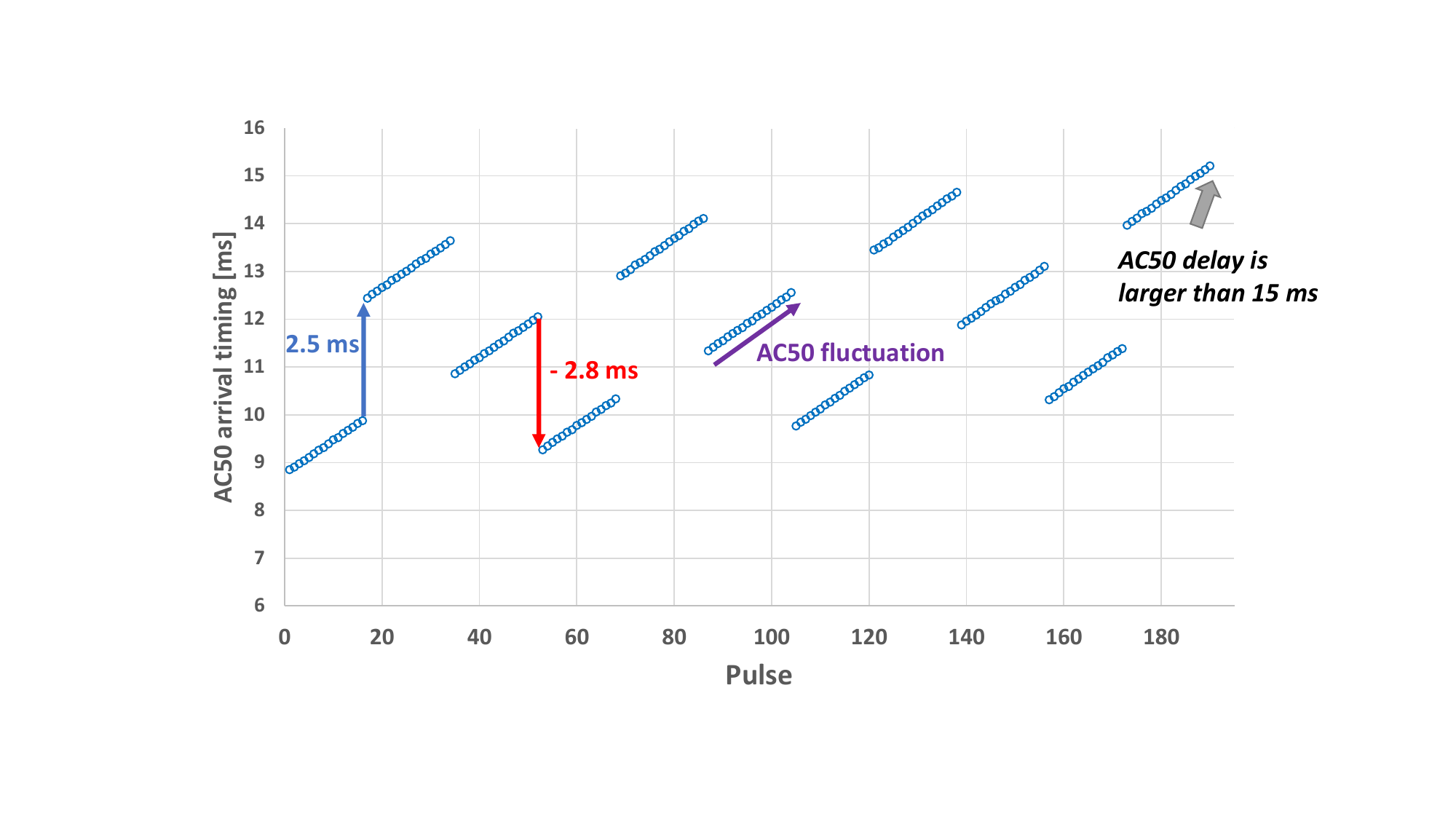}
    \caption{An example of 16/18-pulse sequence shift measured by TDC. The AC50 arrival time becomes late than 15 ms when AC50 fluctuates strongly.}
    \label{fig:ac_drift}
\end{figure}

Figure~\ref{fig:ac_drift} shows the change of the AC50 arrival timing measured by TDC under 16/18-pulse sequence shift. The data is taken from the timing system log in 2020. The slope inside each sequence indicates the value of $\Delta_{AC}$, which is $\sim$70 $\mu$s on average in Fig.~\ref{fig:ac_drift}. The initial AC50 arrival timing is $\sim$9 ms and it becomes larger than 15 ms after several sequences. The 16/18-pulse sequence shift fails to keep the AC50 arrival timing be close to the $T_{ref}$, resulting in a race condition.

\subsection{Simulation of Sequence Shift}
\label{sub:simu_seq_shift}

\begin{figure}[!hbt]
    \centering
    \begin{minipage}[c]{8cm}
        \centering
        \includegraphics[width=2.4in,height=2.2in]{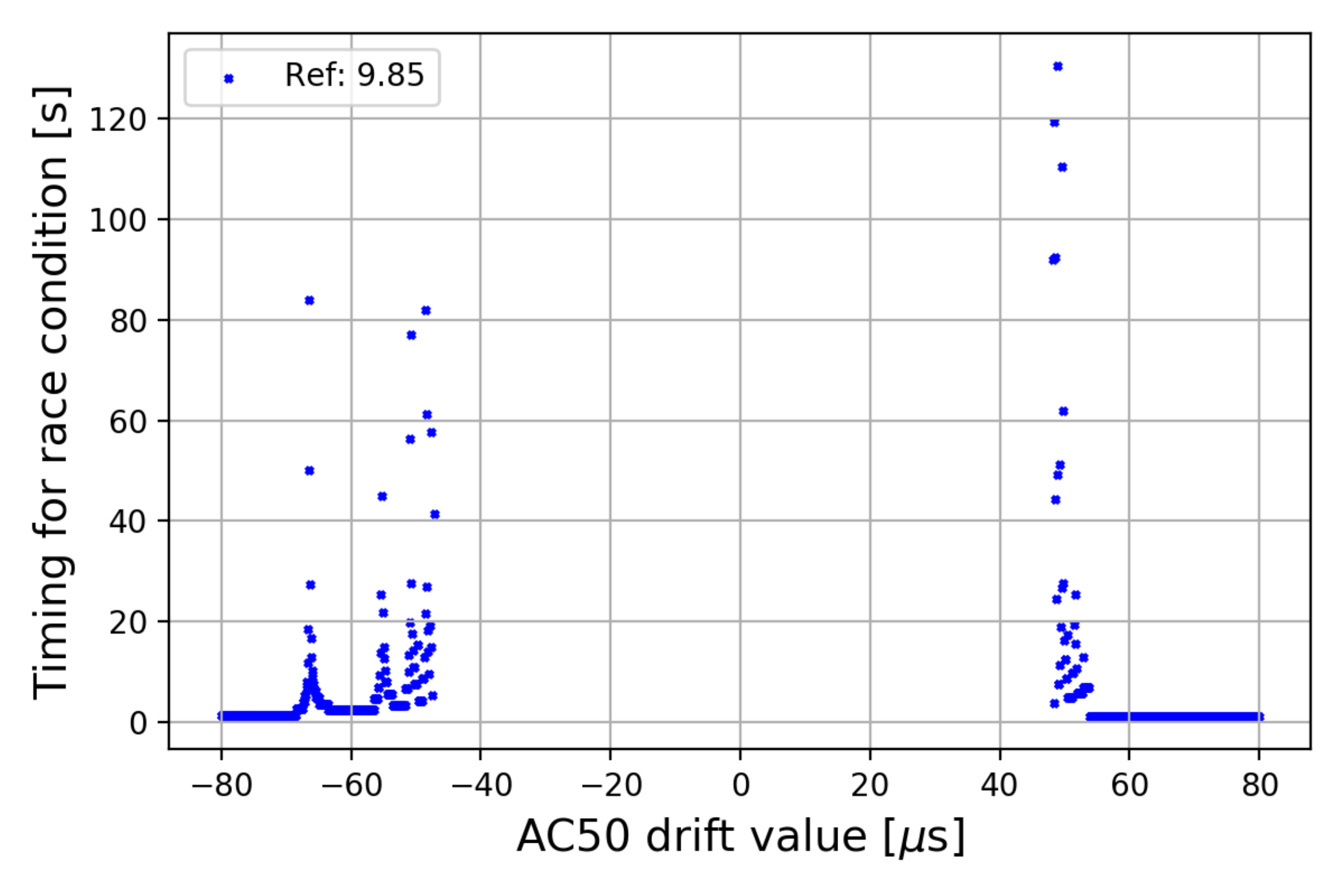}
        \caption{How much time it take to enter the race condition for the 16/18-pulse sequence shift.}
        \label{fig:ac_drift_simulation_16_18}
    \end{minipage}
    \begin{minipage}[c]{8cm}
        \centering
        \includegraphics[width=2.5in,height=2.2in]{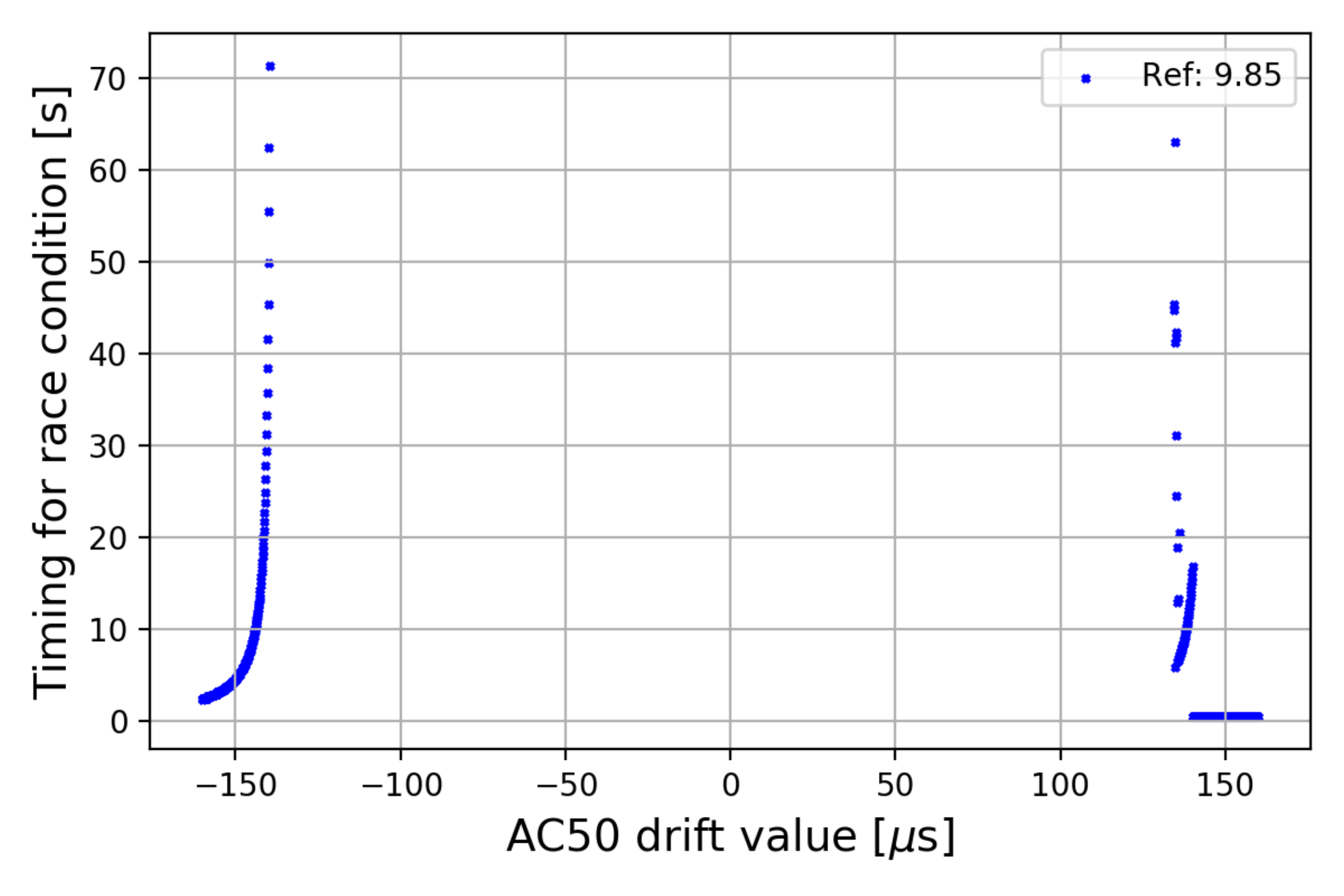}
        \caption{How much time it take to into the race condition for the 8/9-pulse sequence shift.}
        \label{fig:ac_drift_simulation_8_9}
    \end{minipage}
\end{figure}



To figure out the relation between the $\Delta_{AC}$ and race condition, a simulation was performed according to the sequence shift algorithm. We simulated the process of the sequence shift by simply assuming a fixed $\Delta_{AC}$. The fastest timing for the AC50 arrival timing to enter the race condition is examined and recorded. For example, if $\Delta_{AC}$ is 0, meaning that the AC line frequency is ideal 50 Hz, the sequence shift algorithm selects the sequence type in a fixed order and the race condition never happens.

Figure~\ref{fig:ac_drift_simulation_16_18} shows the simulation result of 16/18-pulse sequence shift. The vertical axis represents the shortest time it takes to enter the race condition under a certain AC50 drift value. The initial AC50 arrival timing is 10 ms, and $T_{ref}$ is 9.85 ms. If $|\Delta_{AC}| <= 40\mu$s, the 16/18-pulse sequence shift can always compensate for the AC50 drift, and race conditions will not happen. While $\Delta_{AC}$ is $\sim$45 $\mu$s, the timing system goes into race condition after $\sim$130 s. If $\Delta_{AC}$ becomes larger, e.g., 60 $\mu$s, the timing for race condition only requires $\sim$0.6 s. Note that the figure is asymmetric because both the sequence shift value (2.5 and 2.8 ms) and boundary for race conditions (4.5 and 15 ms) are different.

Similarly, the tolerance of the AC50 drift for the 8/9-pulse sequence is simulated in Fig.~\ref{fig:ac_drift_simulation_8_9}. The tolerance is enlarged to approximately 120 $\mu$s as the compensation opportunity for AC50 drift becomes twice larger.

\section{Stabilization of timing system} 
\label{sec:error_handling}
To guarantee a stable timing system and avoid the race condition, the following solutions are discussed.

\subsection{AC50 Independent Operation}
The reason why the timing trigger is synchronized with AC line frequency is described in Section~\ref{sub:ac_sync}. However, the original experiment data about the beam energy jitter variance under different AC line phases was lost due to historical reasons. Besides, many recent power supply systems can operate independently of the AC line frequency, the timing system does not need to synchronize with the AC line frequency. The timing system at J-PARC is an example~\cite{2003tamuraJPARCTimingSystem}. Some devices in LINAC are also replaced in the past. Thus, an experiment is performed to evaluate the effect of AC50-independent operation at present.

\begin{figure}[!hbt]
    \centering
    \includegraphics*[width=.8\columnwidth,height=3.5in]{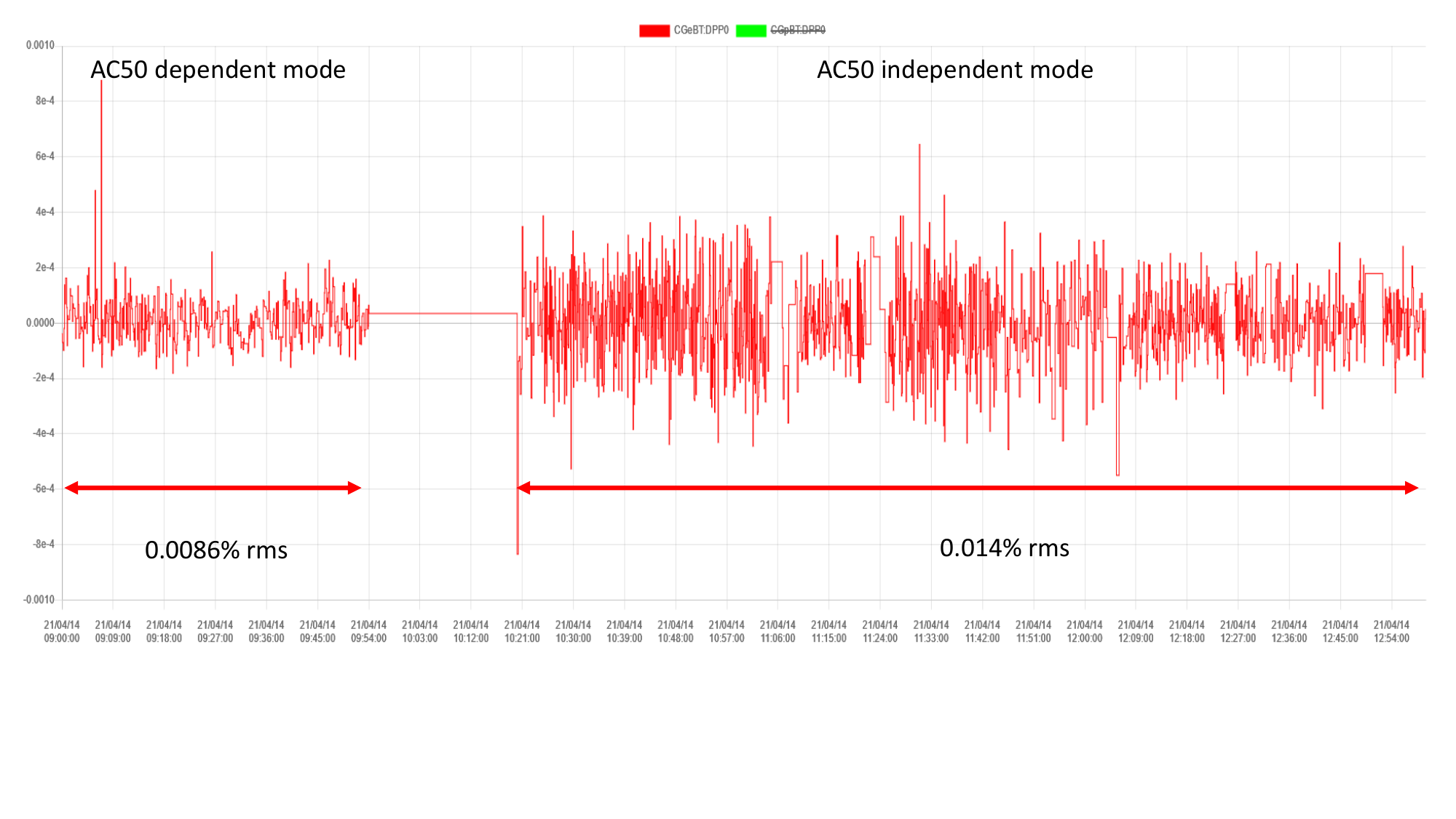}
    \caption{The average energy jitter measured at the beam transport line of LINAC.}
    \label{fig:energy_ac50}
\end{figure}

After replacing the MTG module with a strictly 50 Hz signal generator, the timing trigger signal is delivered at a different phase of the AC line. The hardware at LINAC works under the AC50-independent mode. The energy jitter of electron bunch at the beam transport line of LINAC is measured and the results are displayed in Fig.~\ref{fig:energy_ac50}. The average energy jitter for AC50-dependent mode is 0.0086\% (r.m.s) and it increases to 0.014\% (r.m.s) after switching to the AC50-independent mode. Though the beam quality is indeed related to AC line frequency, the overall energy jitter is still acceptable for SuperKEKB MR as the required energy spread in the current operation stage is 0.1\%~\cite{2018furukawaRejuvenation7GevSuperKEKB}. Consequently, the injection for SuperKEKB rings can be performed without synchronizing to the AC line.

During the experiment, a strong dependency between AC line frequency and the power supply of the injection kicker magnets at PF is observed. The timing trigger for PF ring injection must synchronize with the AC power line.

\subsection{AC50 regulator}

\begin{figure}[!hbt]
    \centering
    \includegraphics*[width=.9\columnwidth]{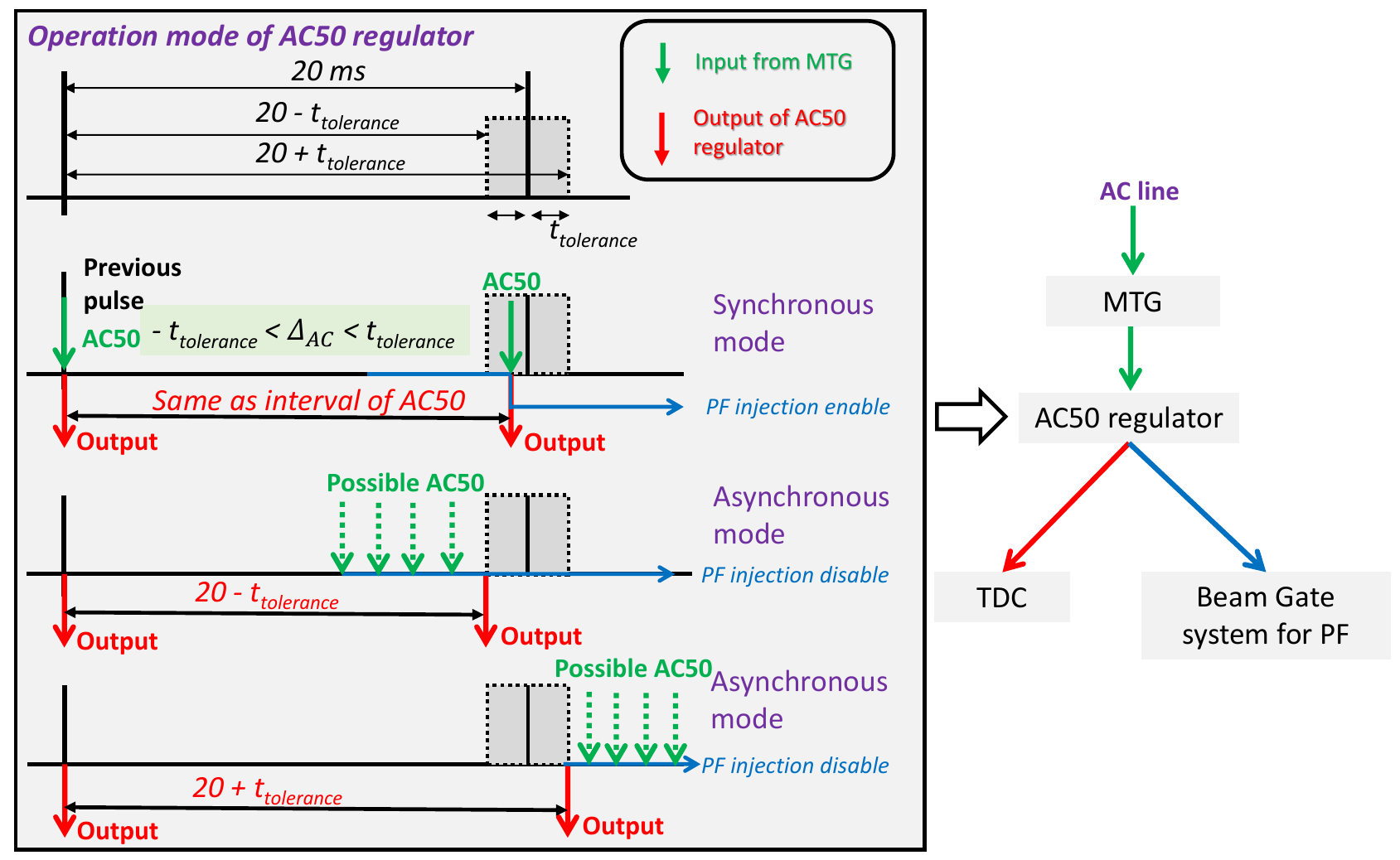}
    \caption{Schematic diagram of the AC50 regulator module. If AC50 fluctuates strongly, the AC50 regulator enters asynchronous mode to keep the sequence shift stable.}
    \label{fig:ac50_regulator}
\end{figure}


Considering that strong AC50 drift occurrence is rather rare according to the data measured by TDC (see Fig.~\ref{fig:ac_drift_hist}), a CompactRIO-based module called AC50 regulator is designed to switch between the AC50-dependent and AC50-independent modes. As Fig.~\ref{fig:ac50_regulator} shows, the AC50 regulator module is inserted between MTG and TDC modules. It supports two operation modes, synchronous mode and asynchronous mode, which correspond to AC50-dependent and AC50-independent operation, respectively. Under synchronous mode, the AC50 regulator module directly passes the AC50 signal to the TDC module. On the other hand, the module generates outputs with a fixed interval to TDC when the AC50 drift value is large. And the timing system starts to operate under AC50-independent mode in a short period. When the AC50 drift value gradually becomes smaller, the module switches to synchronous mode within several seconds, and the timing system trigger synchronizes with the AC line again.

A parameter called $t_{tolerance}$, which is decided based on the simulation result of sequence shift (see Fig.~\ref{fig:ac_drift_simulation_16_18}), is used to determine the maximal AC50 drift value that the AC50 regulator permits. If AC50 drift value $|\Delta_{AC}|$ is larger than $t_{tolerance}$, the module enters asynchronous mode. The range of output interval is restricted to be $20-t_{tolerance}$ or $20+t_{tolerance}$. Meanwhile, the PF injection is disabled by sending a signal to the beam gate system. Combining the simulation results and data analysis, the value of $t_{tolerance}$ is set to 40 $\mu$s at LINAC. The output interval of the AC50 regulator is within 20$\pm$0.04 ms. About 1\% pulses are prohibited for PF injection.

\subsection{Performance of AC50 regulator}


\begin{figure}[!hbt]
    \centering
    \includegraphics*[width=.6\columnwidth]{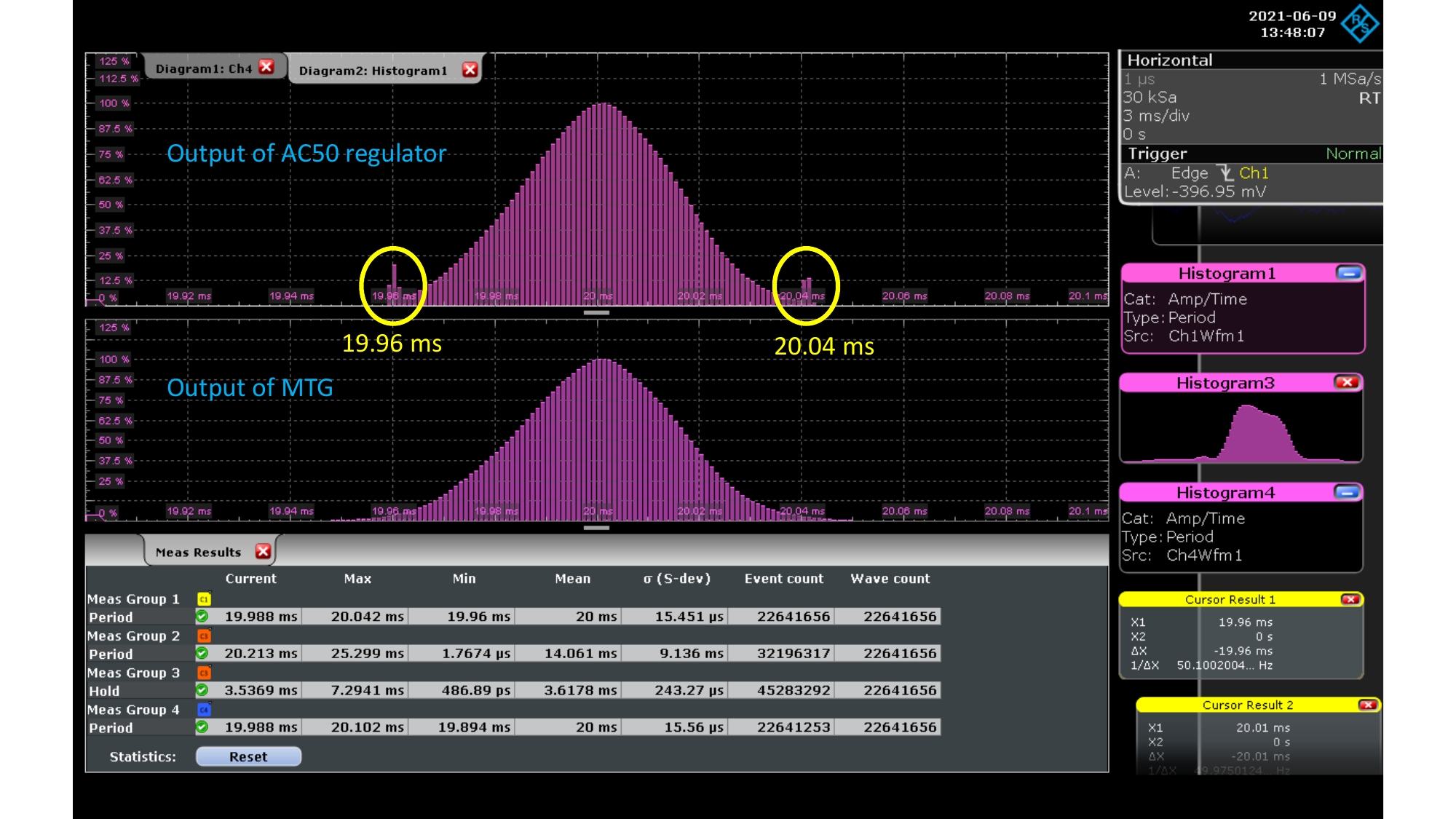}
    \caption{Comparison of AC50 regulator and MTG output. The period of AC50 regulator output restricts to 20$\pm$0.04 ms.}
    \label{fig:ac_regulator_hist}
\end{figure}

The AC50 regulator module is installed and tested in the 2021 spring. Since then, the timing system failure caused by strong AC50 fluctuation never happens. The processing delay of the AC50 regulator is $\sim$500 ns. The long-term stability of the AC50 regulator module is studied by measuring the output signal interval of the AC50 regulator and MTG module. The histogram shown in Fig.~\ref{fig:ac_regulator_hist} indicates that the range of the AC50 regulator output is guaranteed within 20$\pm$0.04 ms.

\begin{figure}[!hbt]
    \centering
    \includegraphics*[width=.6\columnwidth,height=3in]{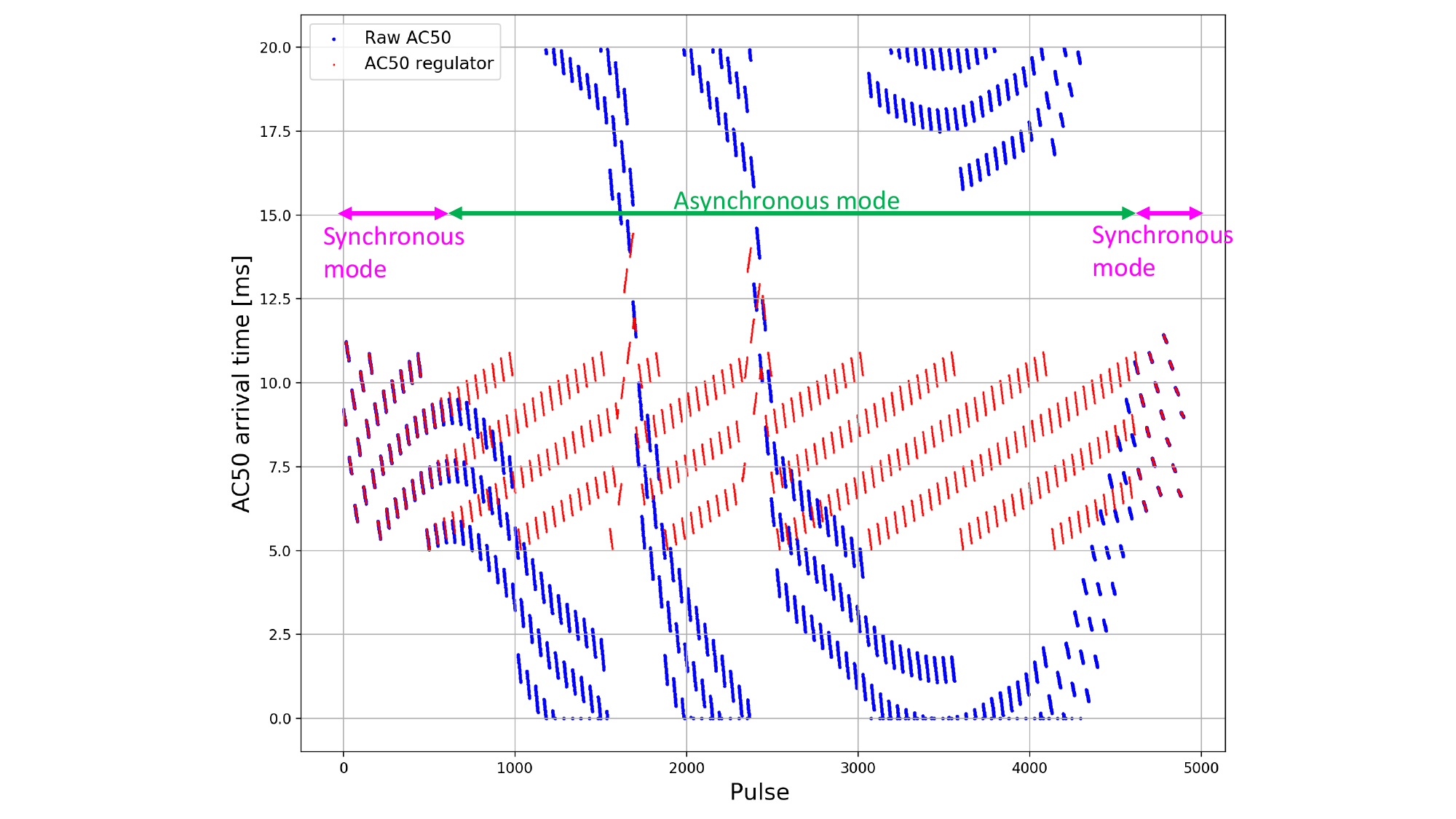}
    \caption{The output of AC50 regulator resynchronizes to AC50 after strong AC50 fluctuation stops.}
    \label{fig:ac_recovery}
\end{figure}
The AC50 regulator must switch back to synchronous mode after strong AC drift stops. There are only theoretical possibilities that the frequencies of AC line and regulator output are the same, though remaining at a different phase. In practice, the AC line always fluctuates, and it will catch up with regulator output after a short period. According to the data measured by TDC in one month, the average transition time from asynchronous mode to synchronous mode is $\sim$31 s. Figure~\ref{fig:ac_recovery} displays an example of the transition process. The red dots are the arrival time of regulator output while the blue dots are the arrival time of AC50. The asynchronous mode lasts for $\sim$80 s and it takes $\sim$20 s to resynchronize with the AC line after strong AC drift stops.

\subsection{Improvement for sequence shift}
The AC50 regulator stabilizes the timing system and reduces the accelerator failure time at LINAC. However, there are still concerns about the degradation of the beam quality. To operate the timing system under AC50-dependent mode, two sequence shift improvements are proposed to optimize the timing system.
\begin{figure}[!hbt]
    \centering
    \includegraphics*[width=.8\columnwidth]{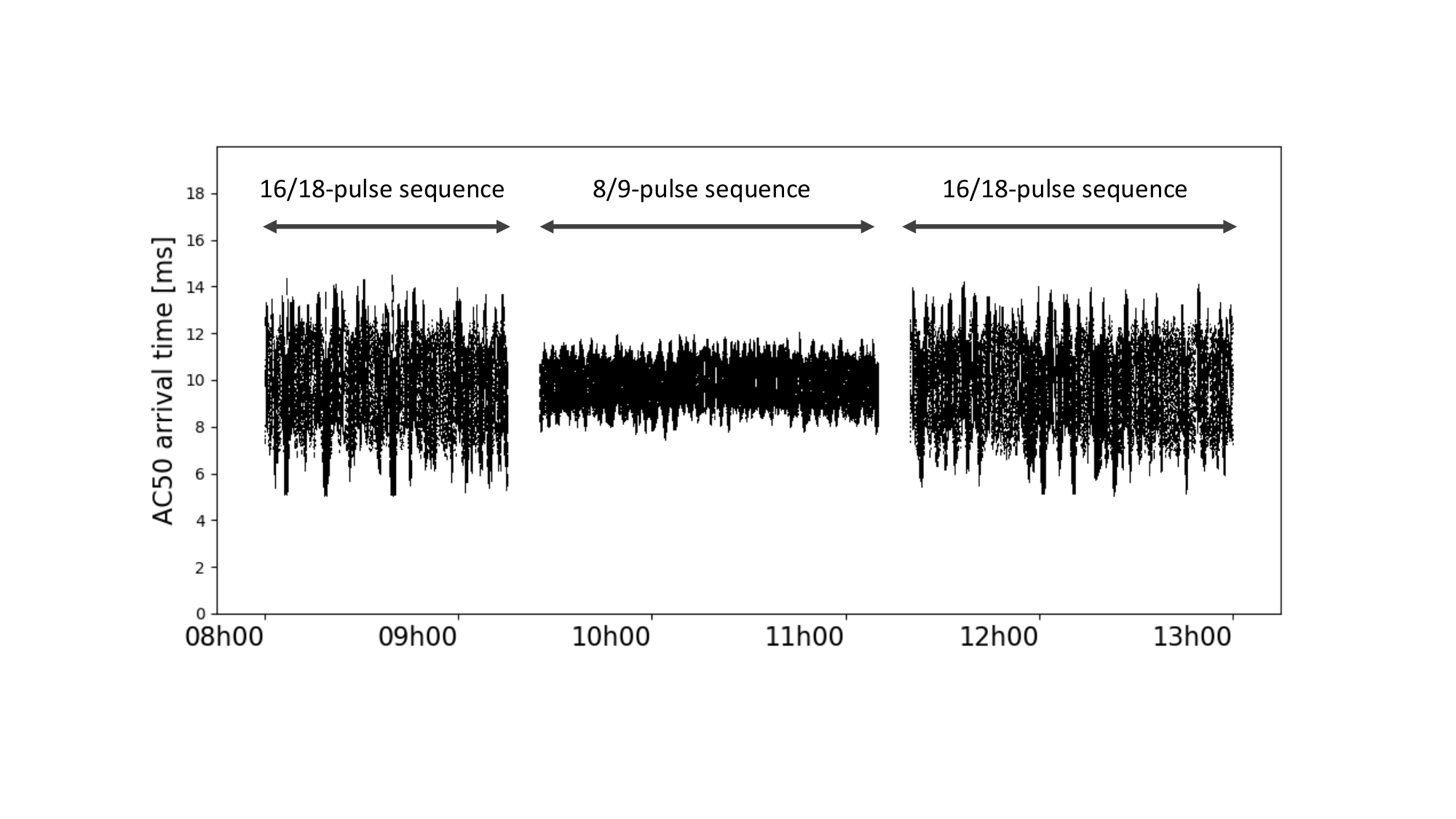}
    \caption{Comparison of AC50 arrival timing between 16/18-pulse sequence and 8/9-pulse sequence.}
    \label{fig:comparison_sequence_shift}
\end{figure}

As Fig.~\ref{fig:comparison_sequence_shift} shows, compared with the 16/18-pulse injection sequence, the 8/9-pulse injection sequence increases the robustness of the sequence shift algorithm. Switching to the 8/9-pulse injection sequence is a possible method to solve the timing failure because the AC50 adjustment speed is faster. The restriction of the sequence length originates from the DR storage time and the root reason is the coupling of DR injection and extraction calculation. The DR extraction delay is calculated at the injection pulse to keep the relation shown in Table~\ref{table:bs_delay}. Shifting the RF phase at downstream LINAC is an efficient way to increase the DR extraction opportunity and relieve the stress of long sequence management. The DR extraction delay is calculated based on the shifted RF phase at downstream LINAC. Under these circumstances, the dependency between DR and MR injections can be removed. The operation of 8/9-pulse sequence shift is possible as the DR injection and extraction delay calculations can be separated. More descriptions about RF phase shifting can be found in~\cite{2018kajiBUCKETSELECTIONSuperKEKB}.

\begin{figure}[!hbt]
    \centering
    \begin{minipage}[c]{8cm}
        \centering
        \includegraphics[width=2.6in,height=2.4in]{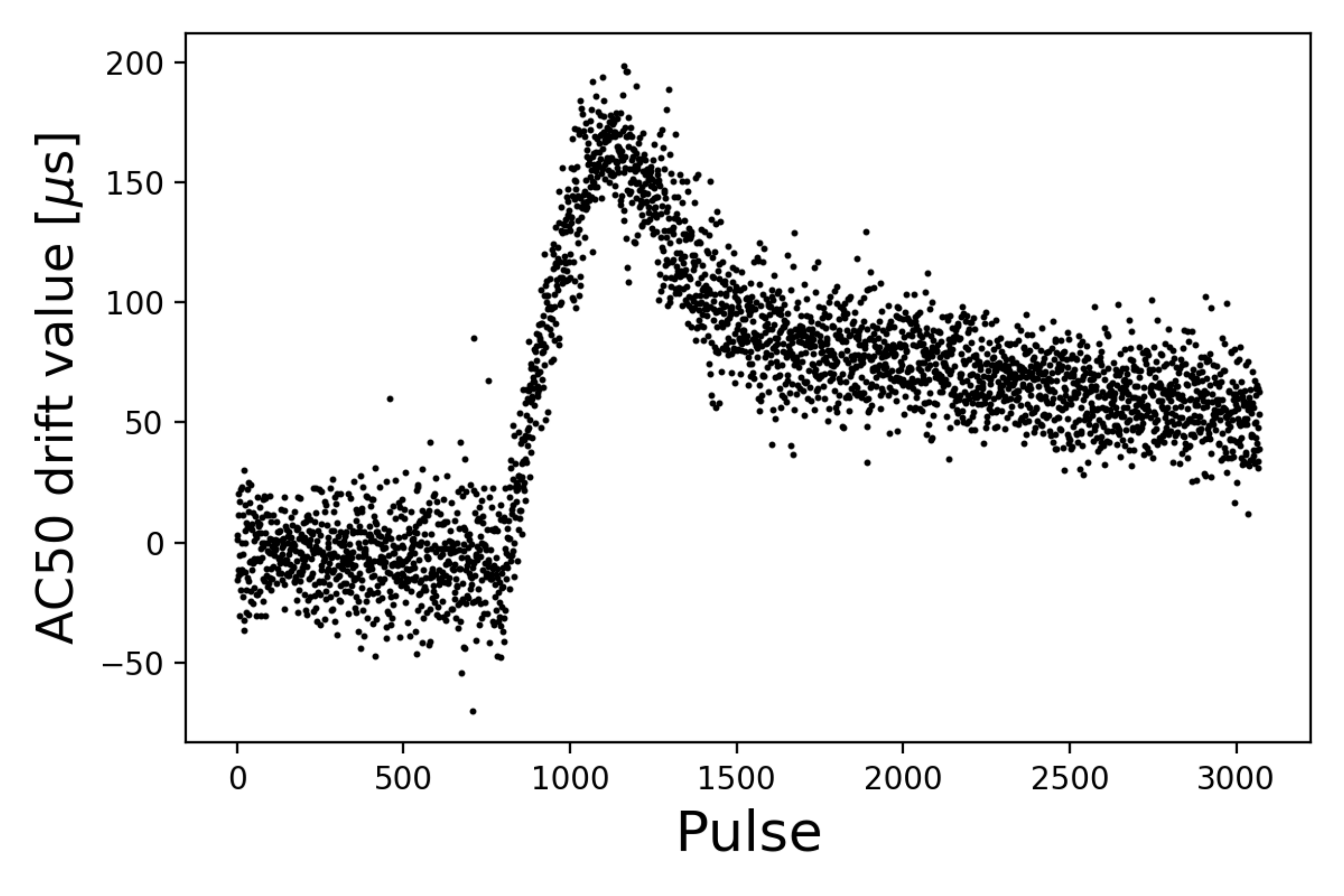}
        \caption{Extremely strong AC50 drift recorded by TDC.}
        \label{fig:ac_drift_20191012}
    \end{minipage}
    \begin{minipage}[c]{8cm}
        \centering
        \includegraphics[width=2.8in,height=2.4in]{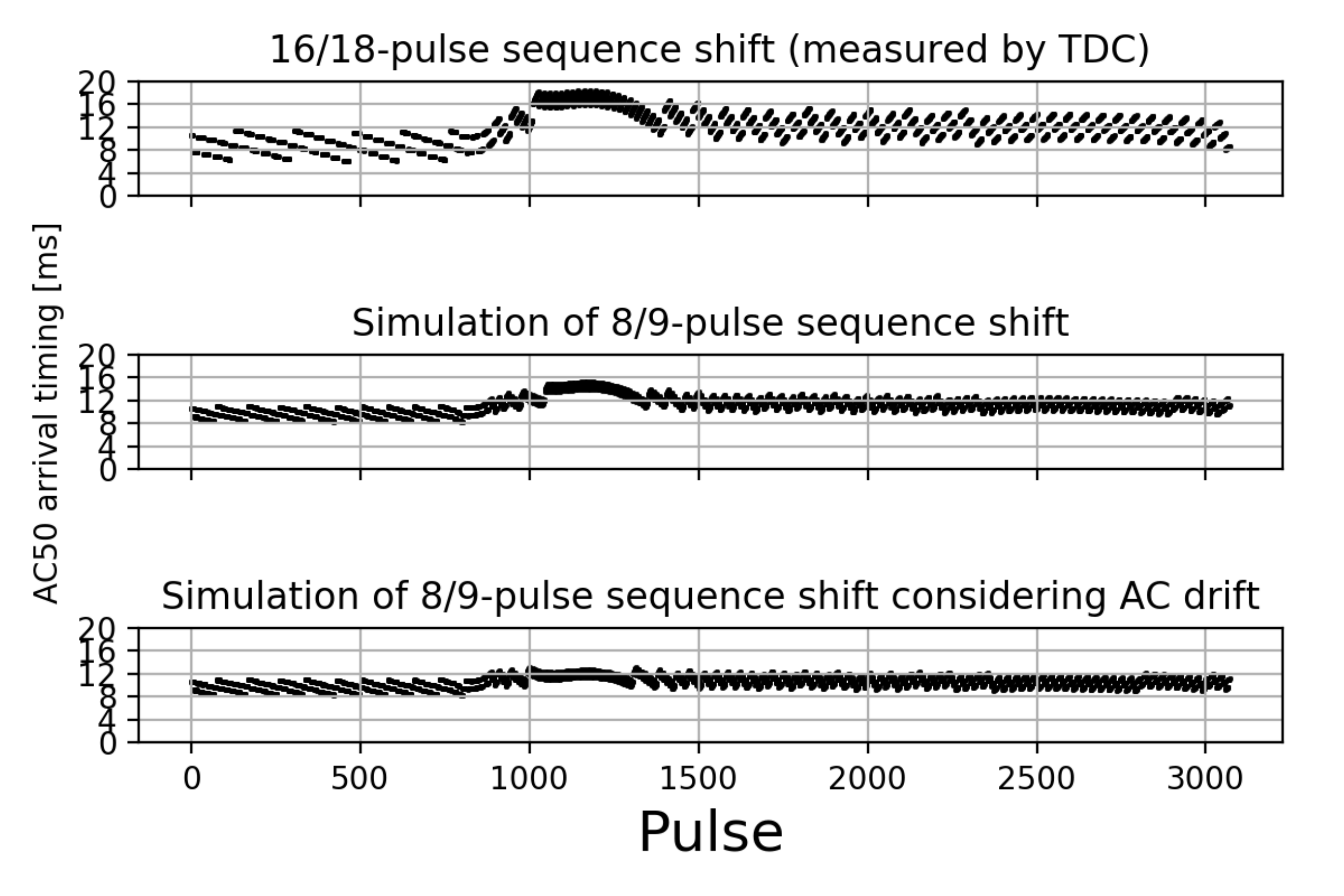}
        \caption{Comparison of different sequence shift schemes.}
        \label{fig:compare_seq_shift}
    \end{minipage}
\end{figure}

Apart from the 8/9-pulse sequence shift, the sequence shift algorithm can be further improved by considering the AC50 drift value. Currently, the estimated AC50 arrival time is used to determine the next sequence type. If strong AC50 drift happens, the estimated value has a large deviation from the real value. The timing system could enter race conditions when an inappropriate sequence type is selected. To avoid such a situation, the sequence shift algorithm is modified and the average AC50 drift value in recent pulses is utilized to help the sequence shift algorithm to estimate the AC50 arrival time. The estimated AC50 value becomes closer to the real value even when AC50 fluctuates strongly. According to our simulation, the new algorithm can handle an AC50 drift ranging from -156 $\mu$s to 158 $\mu$s. Figure~\ref{fig:ac_drift_20191012} shows a timing system failure which is caused by the strongest AC50 drift situation that we have recorded during the past two years. The data was recorded by TDC in 2019. The AC50 drift value reached 120 $\mu$s and lasted for several seconds. Figure~\ref{fig:compare_seq_shift} compares three sequence shift schemes, 16/18-pulse sequence shift which is currently using, 8/9-pulse sequence shift, and 8/9 pulse sequence shift with the functionality of using the past AC drift value. The results show that the new algorithm can handle the most significant AC50 drift situation.

The 8/9-pulse sequence operation with AC50 drift compensation has been planned for the 2021 summer.

\section{Summary}

This article provides a systematic analysis of the timing system and introduces the basic concept of a timing system, such as the event system, bucket selection, and RF synchronization. The timing system discussed here is strongly associated with other accelerator subsystems. Several restrictions which originate from the DR operation and AC line significantly increase the complexity of the timing system. The robustness of the timing system should be emphasized to achieve a stable beam operation.

The timing system failures that happened at LINAC frequently interrupt the operation of SuperKEKB. We analyzed the algorithm of the injection sequence shift and used a simulation to identify the relation between the AC50 fluctuation and the timing system failure. After a beam quality experiment, we found that it is acceptable for SuperKEKB to operate the timing system under the AC50-independent mode. Therefore, an AC50 regulator module is designed to switch between AC50-independent and AC50-dependent modes. The performance and stability of the new module are tested. The timing system failures caused by strong AC50 fluctuation are solved after deploying the AC50 regulator module. Furthermore, to avoid the beam quality degradation caused by AC50-independent operation, we also propose to upgrade the sequence shift by applying short sequence length and performing a reliable AC50 estimation. The simulation results show that the new method can handle the most severe AC drift situation we have ever met in the last two years. The analysis procedure and solution for a timing system with a long bucket selection cycle can be referred to for the design of future circular accelerators.


\section{Declaration of competing interests}
The authors declare that they have no known competing financial interests or personal relationships that could have appeared to influence the work reported in this paper.

\section{Acknowledgments}
The author is grateful for the scholarship provided by China Scholarship Council (No.201804910435).

\bibliography{timing}


\end{document}